\documentclass[%
 reprint,
 amsmath,amssymb,
 aps,
]{revtex4-1}

\usepackage{graphicx}
\usepackage{dcolumn}
\usepackage{bm}

\usepackage{amsmath}
\usepackage[export]{adjustbox}
\usepackage{enumitem}
\usepackage[usenames, dvipsnames]{color}
\usepackage{relsize}
\usepackage{enumitem}
\usepackage{lipsum}
\usepackage{varwidth}

\begin{document}

\preprint{APS/123-QED}

\title{Splitting Nodes and Linking Channels: A Method for Assembling Biocircuits from Stochastic Elementary Units}

\author{Cameron Ferwerda}%
\affiliation{%
 School of Physics and Astronomy, University of St Andrews, North Haugh, KY16 9SS, Scotland UK}
 \author{Ovidiu Lipan}%
 \email{olipan@richmond.edu}
\affiliation{%
 Department of Physics, University of Richmond, 28 Westhampton Way, Richmond VA, 23173
}%

\date{\today}

\begin{abstract}
Akin to electric circuits, we construct biocircuits that are manipulated by cutting and assembling channels through which stochastic information flows. This diagrammatic manipulation allows us to create a method which constructs networks by joining building blocks selected so that (a) they cover only basic processes; (b) it is scalable to large networks; (c) the mean and variance-covariance from the Pauli master equation form a closed system and; (d) given the initial probability distribution, no special boundary conditions are necessary to solve the master equation. The method aims to help with both designing new synthetic signalling pathways and quantifying naturally existing regulatory networks.
\begin{description}
\item[PACS numbers]
87.18.Cf., 87.18.Nq., 87.18.Tt.
\end{description}
\end{abstract}

\maketitle


\section{\label{sec:level1}Introduction}

Networks of bio-molecular pathways orchestrate the development, progress and fate of living cells.
Currently there is a struggle to translate the experimental results into pictorial representations of molecular signalling pathways \cite{Raza2008}. These pictorial representations are necessary for understanding biological processes at a systems level as used in everything from drug discovery to classification of biological processes. As the networks and processes grow more complex, the need for computation becomes apparent because extensive textual explanation of pictorial representation of information flow through pathways containing hundreds of molecules is inefficient and impractical.

In this paper we use stochastic computation because signals that propagate through successive molecular events are stochastic in nature. Genetic regulatory reactions involve a range of molecule numbers from the thousands down to singular molecules. The statistical fluctuations at low molecule numbers are usually higher relative to the mean values and thus have a strong impact on the cell fate \cite{Elowitz}. Some pathways evolved to use these fluctuations to cell's advantage for driving the cell into diverse phenotypic outcomes \cite{ArkinStochastic}. Phenotypic diversity of an isogenic population caused by stochastic fluctuations is commonly found in microorganisms' response to stress and virulence factors \cite{NoisyBusiness}.

Stochastic fluctuations are often studied by simulating a whole array of stochastic paths for the dynamics of the system. From these stochastic paths the mean values, the standard deviations, and the correlation functions are then computed. This approach quickly becomes impractical for large networks as they are computationally expensive.

Instead of first generating a whole array of stochastic data, the method presented in this paper produces means and the variance-covariance matrix from the Pauli master equation. Many methods of computation \cite{4,5,6,7,8} based on the Pauli master equation \cite{9,10,11,12} have been used to describe the molecular events and their mutual dependence. The master equation is valid for any range of molecular number, from very large for some species to very small for others. However, with the exception of a few simple models the master equation is very difficult to solve. The main reason is that it delivers an infinite system of equations for the moments of the probability distribution. For the past 60 years, moment closure methods have been used to tackle the master equation by reducing the system of equations to make it finite \cite{13,14,15,16,17,18,19,20,26}. The approximation which reduces the equations, known as moment closure, is carried out in a variety of ways. The moment closure method in \cite{SinghHespanha} is achieved by matching time derivatives at an initial time. The resulting Taylor series argument reveals that the time trajectories remain closed for short time intervals. Multiplicative, rather than additive, moments are introduced in \cite{Keeling} and the approximation is made by setting the third order multiplicative moments equal to 1. The  model in \cite{Verghese} assumes that the central moments of third-order are negligible. Approximation in \cite{Rogers} is achieved by entropy maximization under known constraints to avoid unmotivated bias. In \cite{Bradley} techniques and benchmark models are used to compare the different moment closure techniques like mean-field, normal closer, min-normal closure, log-normal closure.

These methods tend to focus on disentangling the equations without considering the topology of the biocircuit. By keeping the topology of the biocircuit in the forefront the method presented here uses the diagrams themselves to implement the moment closure. Because each term in the master equation has a unique pictorial representation, there is a simple correspondence between the qualitative interactions depicted by the biochemical pathway and the mathematical model. This gives a method that is diagrammatically easy to use and manipulate by researchers not interested in the numerical details, but also retains all of the quantitative properties of the master equation that are useful for extensive computation.

In what follows we describe the method through which the channels and nodes are split and later rejoined to create moments that close at second order by using the ubiquitous equilibrium reaction, $A+B\leftrightarrows C$ (Sec. II). The complex formation and its reverse process, the dissociation, are the most important elementary reactions. For example, irreversible complex formation is key to DNA error correcting and T-cell recognition \cite{Hopfield, Goldstein}. Then we identify a set of three elementary units, which together with $A+B\leftrightarrows C$, are used to construct signalling pathways (Sec. III). We show how our method can be used for two important types of networks: Bistable (Sec. IV) and Ultrasensitive (Sec. V). Finally, we explore the concept of modularity by splitting nodes and projecting the large circuit into smaller circuits using the elementary units (Sec. VI).

\section{\label{sec:Section1} Splitting the nodes and linking the channels}

Signals processed by a network composed of $N$-molecule types consist of stochastic time-dependent levels of molecular numbers, $q=(q_k)$, $k=1,\dots,N$. The environmental inputs and the way the molecules control themselves is described by the set of transition probabilities per unit time $T_{\epsilon}(q,t)$. Molecules can jump from one state $q$ to another $q+\epsilon=(q_k+\epsilon_k)$, where $\epsilon$ is an N-vector given by stoichiometry with $\epsilon_k \in\{ \pm1,0\}$ representing the jumps that either increase, decrease or do not change the molecule number $q_k$.
The Pauli master equation

\begin{equation}\label{MEq}
 \resizebox{0.19\hsize}{!}{$
 \frac{\partial P(q,t)}{\partial t}=
 $}
 \resizebox{0.7173\hsize}{!}{$
   \sum_{\epsilon}T_{\epsilon}(q-\epsilon,t)P(q-\epsilon,t)-P(q,t)\sum_{\epsilon}T_{\epsilon}(q,t)
 $}
\end{equation}
expresses this time evolution of the network.

The first order moments are generated from $F(z,t)=\sum_{q_1=0,...,q_N=0}^{\infty}z_1^{q_1}...z_N^{q_N} P(q_1,...,q_N,t)$ and the second order factorial moments are generated from
$F_{k}=\partial_{z_k} F|_{z=1}$ and $F_{jk}=\partial_{z_j,z_k} F|_{z=1}$, where  $z\equiv(z_1,...,z_N)$. For ease we will refer to factorial moments as moments.

\begin{figure}[h]
\includegraphics[scale=0.25]{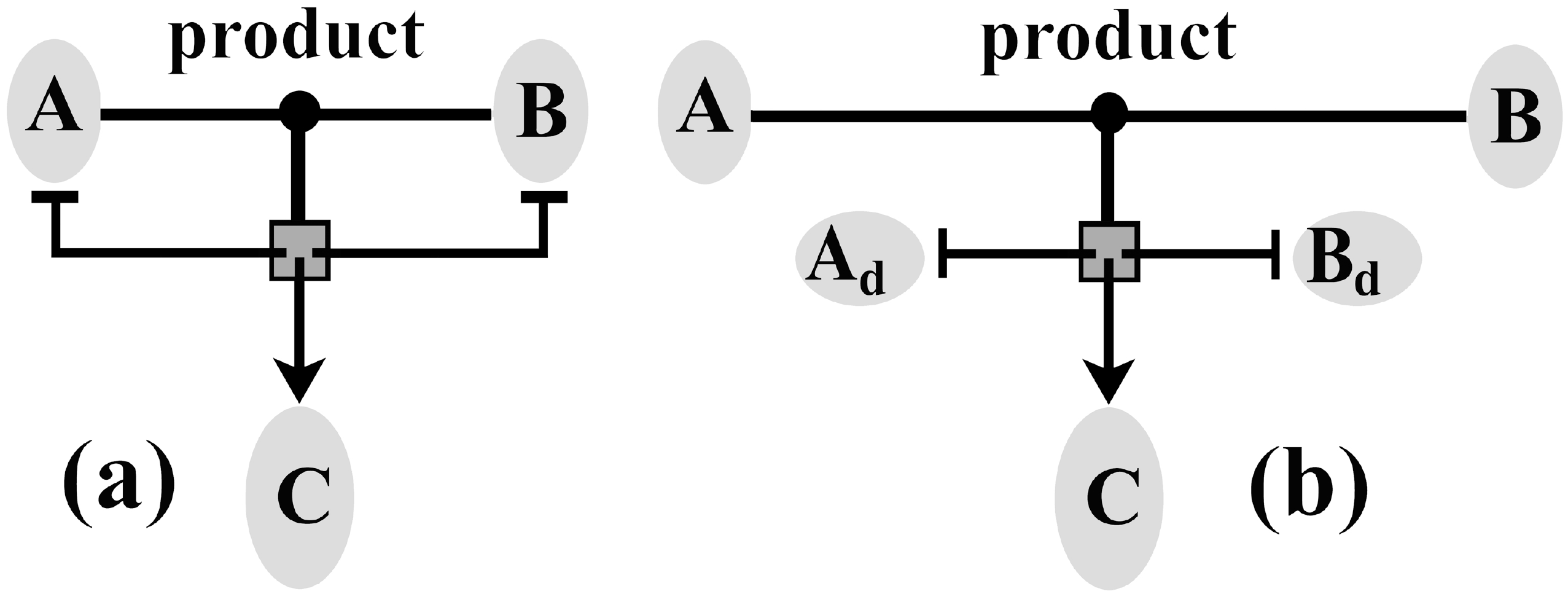}
\caption{\label{fig:Equilibrium}  A and B bind together to form C. (a) The control lines that originate on molecules A and B meet at the node that symbolizes the product $q_A q_B$ and end on the box. The box depicts the action $\epsilon=(-1,-1,+1)$. Two action lines start on the box and end on $A$ and $B$ respectively with the end bar denoting the annihilation ($-1$ in $\epsilon$) of those molecules. The third action line ends in an arrow on C, expressing the creation process ($+1$ in $\epsilon$). (b) The open-loop biocircuit.  }
\end{figure}

The first basic building block is the irreversible complex formation, where molecule A binds to molecule B with  $T(q,t)=k q_A q_B$ to form the complex C, represented in Fig.\ref{fig:Equilibrium}(a) as a control-action diagram \cite{LipanIEE, LipanModernPhys}.

The master equation for $F(z_A,z_B,z_C,t)$ is
\begin{equation}\label{eq:ProductTerm}
 \begin{split}
 \partial_t F= (z_{A}^{-1}z_{B}^{-1}z_{C}^{+1}-1) k\, z_{A}z_{B}\partial_{z_A z_B} F.
 \end{split}
\end{equation}

\begin{figure}[h]
\includegraphics[scale=0.6]{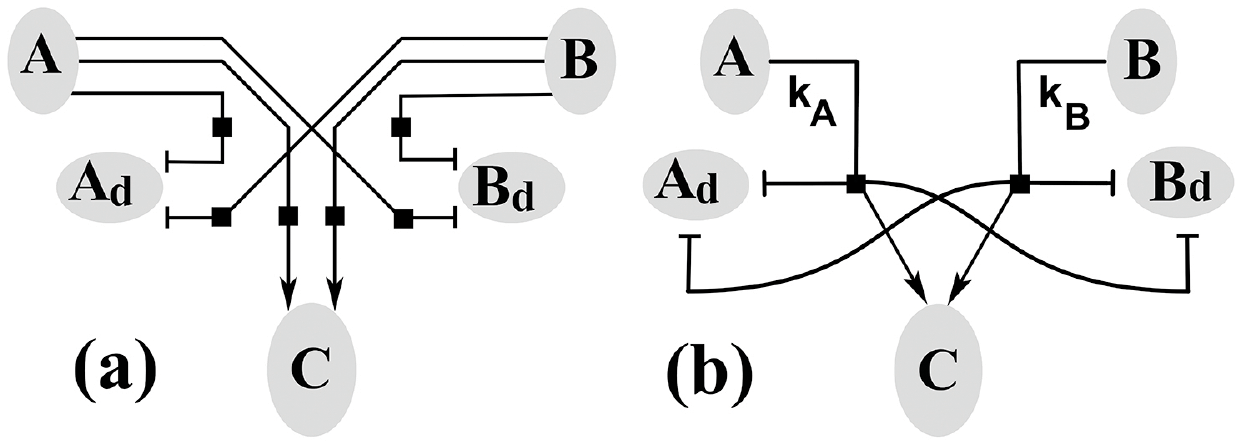}
\caption{\label{fig:SplitProductBoth} The product node from Fig.\ref{fig:Equilibrium}(b) is split into either six channels (a) or two channels (b). }
\end{figure}

The problem we face is that the time evolution of Fig.\ref{fig:Equilibrium}(a) never closes at any moment order due to the complex formation product $q_A q_B$ which gives the second derivative in (\ref{eq:ProductTerm}). For example, applying $\partial_{z_A z_B}$ on (\ref{eq:ProductTerm}), the time evolution obtained for the second order moment $F_{AB}(t)$ turns out to be dependent on third-order moments. To obtain a closed stochastic model we propose an approach which is based on the interpretation of the diagram from Fig.\ref{fig:Equilibrium}(a) as not only a place-holder for the interactions, but as a more literal flow of information through the biocircuit. In Fig.\ref{fig:Equilibrium}(a) the information that flows from $A$ and $B$ is multiplied at the 'product' node. Then, after it passes through the action node (the square-shaped node), it flows into $C$ and feeds back to $A$ and $B$. This feedback prevents moment closure. To obtain a finite system of equations we break the feedback by duplicating molecules $A$ and $B$ into $A_d$ and $B_d$, Fig.\ref{fig:Equilibrium}(b). The open-loop biocircuit is finite and completely solvable, but it closes at fourth-order moments \cite{Narvik}. To reduce it to second order moments we split the product node and the action box to let the information flow from $A$ to $C$ on a different channel than that from $B$ to $C$. There are many ways to produce this splitting. For example, in Figs.\ref{fig:SplitProductBoth}(a) and (b) there are six channels and two channels respectively. We prefer the option from  Fig.\ref{fig:SplitProductBoth} (b) over that from (a) because the requirement that the master equation is free from boundary conditions will be easily enforced on option (b). After splitting the channels, a transition probability must be assigned to each one. To guarantee the closing of the evolution equations at second order, the transition probabilities assigned to each channel in Fig.\ref{fig:SplitProductBoth} (b) will be $k_A q_A$ and $k_B q_B$ for the channels starting from $A$ and $B$ respectively.

\begin{figure}[h]
\includegraphics[scale=0.75]{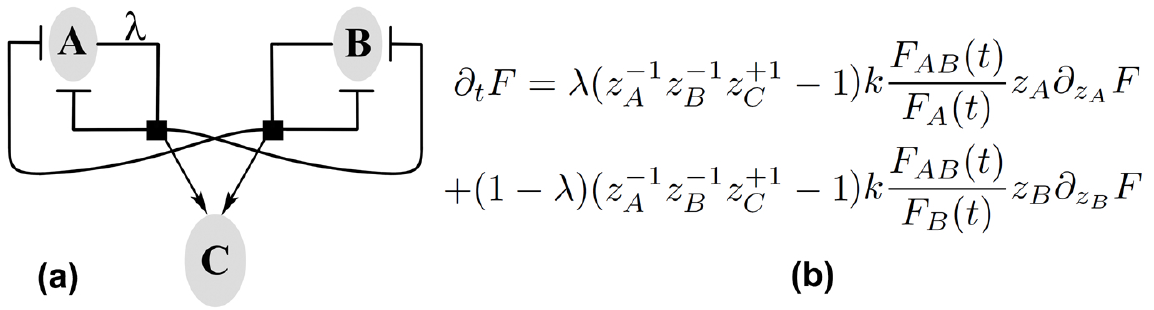}
\caption{\label{fig:SplitProductClosedLoops} (a) The closed-loop biocircuit of Fig.\ref{fig:SplitProductBoth}(b). The duplicated molecules $A_d$ and $B_d$ are glued back to $A$ and $B$ respectively. (b) The LC-master equation for splitting the product node. The driving parameter $\lambda$ is associated with $A$ and $1-\lambda$ with $B$.}
\end{figure}

Since $A_d$ and $B_d$ are copies of $A$ and $B$, a specific time evolution must be imposed on the biocircuit from Fig.\ref{fig:SplitProductBoth}(b). To specify this evolution we start by dividing the time into small equal intervals $\Delta t$. The initial conditions $F_i(0)$ and $F_{ij}(0)$ at $t=0$ are known for Fig.\ref{fig:Equilibrium}(a). These initial conditions are transferred to the molecules from  Fig.\ref{fig:SplitProductBoth}(b) in such a way that the duplicate molecules $A_d$ and $B_d$ have the same initial conditions as $A$ and $B$. During the time interval $[0,\Delta t]$ the values for the molecules $A_d$ and $B_d$ change but $A$ and $B$ did not evolve in time because there is no process that changes their molecule number in  Fig.\ref{fig:SplitProductBoth}(b). Because $A_d$ and $B_d$ are duplicates of $A$ and $B$, the final values $F_i(\Delta t)$ and $F_{ij}(\Delta t)$ for $A_d$ and $B_d$ are passed to $A$ and $B$ as initial conditions for the next time interval $[\Delta t, 2\Delta t]$. The process is then iterated, $A$ and $B$ drive $A_d$ and $B_d$ which in turn produce the updated values for $A$ and $B$ for each time interval. Through this updated iterative procedure Fig.\ref{fig:SplitProductBoth}(b) is closed and taking the limit
$\Delta t\rightarrow 0$ we get Fig.\ref{fig:SplitProductClosedLoops}(a). To determine $k_A$ and $k_B$ we start with

\begin{align}\label{eq:ProductTermSplit}
  \partial_t H=(z_{A_d}^{-1}z_{B_d}^{-1}z_{C}^{+1}-1) &k_A z_{A}\partial_{z_A} H \\ +(z_{A_d}^{-1}z_{B_d}^{-1}z_{C}^{+1}-1)& k_B z_{B}\partial_{z_B} H\nonumber,
\end{align}
where $H$ is the generating function of Fig.\ref{fig:SplitProductBoth}(b). From the same figure we read that $(q_A,q_B,q_{A_d},q_{B_d},q_C)\rightarrow(q_A,q_B,q_{A_d}-1,q_{B_d}-1,q_C+1)$, thus $\epsilon=(0,0,-1,-1,1)$ for both action nodes. This gives the term $(z_{A_d}^{-1}z_{B_d}^{-1}z_{C}^{+1}-1)$. The transition probabilities, $k_A q_A$ and $k_B q_B$, give us $k_A z_{A}\partial_{z_A} H$ and $k_B z_{B}\partial_{z_B} H$ respectively.

From (\ref{eq:ProductTerm}) and (\ref{eq:ProductTermSplit}) we get $dF_{A}/dt=dF_{B}/dt=-dF_{C}/dt=-k F_{AB}$ and $dH_{A_d}/dt=-k_A H_A -k_B H_B$. The equal evolution condition $H_{A_d}=F_{A}$ is fulfilled if $k_A H_A +k_B H_B= k F_{AB}$. The updating process $H_{A}=H_{A_d}$ implies $H_A=F_A$, which gives

\begin{equation}\label{eq:IdenticalDynamicsF}
   k_A F_A +k_B F_B= k F_{AB}.
\end{equation}

A simple solution to (\ref{eq:IdenticalDynamicsF}) would be an equal split drive between $A$ and $B$ so that $k_A F_A =k_B F_B= k F_{AB}$. However, an equal split drive is not necessarily obvious especially given that the molecule numbers $A$ and $B$ may be very different. The unequal split solution $k_A F_A =\lambda k F_{AB}$ and $k_B F_B =(1-\lambda) k F_{AB}$ confers more freedom to the model. For convenience we will call the entire procedure the loop-closing (LC) method.

The LC-master equation for $F(z_A,z_B,z_C,t)$ is presented in Fig.\ref{fig:SplitProductClosedLoops}(b) which describes the time evolution of Fig.\ref{fig:SplitProductClosedLoops}(a). The transition probabilities, which are the ratios  of the correlation over the mean values, are not constant; they change together with the stochastic evolution of the molecule numbers. In general, the LC-method is composed of the following steps: (i) duplicate molecules by breaking selected feedback loops; (ii) split nonlinear nodes; (iii) assign transition probabilities to the new channels; (iv) close the loops by updating the initial values between the duplicate and the original molecules. We note that the equivalence between Fig.\ref{fig:Equilibrium}(a) and Fig.\ref{fig:SplitProductClosedLoops}(a) is based on the equalities $H_{A_d}=F_{A}, H_{B_d}=F_{B}$. Because $F_A$ and $F_B$ are driven by the second moment $F_{AB}$, the procedure explicitly involves the second moments, Fig.\ref{fig:SplitProductClosedLoops}(b). However it does not explicitly involve the third and higher moments which are compressed into the $\lambda$-parameters. The relevance of the $\lambda$-parameters is further discussed in Sec.IV.

Now that we have reduced the irreversible complex formation to second order, we can use it to study the equilibrium complex formation since the master equation term for disassociation of the complex contains only a first order partial derivative $(z_{A}z_{B}z_{C}^{-1}-1)  k_n\,  z_{C}\partial_{z_C} F$. The LC approximation of the irreversible process, along with the linearity of the disassociation, allows us to explore how close the LC-procedure comes to reproducing the stochastically simulated data of the equilibrium process. The LC differential equations were computed with Mathematica \cite{Wolfram} and the time variation of each moment was compared with the corresponding data simulated with the Gillepsie algorithm \cite{GillespieArticle}.

All errors between two functions of time were computed as average of the relative error on a sequence of sampled times. The initial probability distribution was taken to be concentrated at fixed molecule numbers $q_{A0},\; q_{B0},\; q_{C0}$,\; $F(z_A,z_B,z_C,t=~0)=z_A^{q_{A0}}z_B^{q_{B0}}z_C^{q_{C0}}$. The error covered the range $10^{-5}$ to $10^{-1}$, the most common being $10^{-3}$ \cite{suppEquilibriumReaction}. We rescaled the unit of time so that complex formation transition probability is unity $k_p=1$. Then we varied the other parameter $k_n$ between $10^{-4}$ and $10^4$. The initial molecule numbers for each molecule were varied between $0$ and $10^3$ in different combinations.

We found that in order to obtain low errors the parameter $\lambda$ should be either $0$ or $1$. If the initial molecule numbers $q_{A0}$ is less than $q_{B0}$ then $\lambda=1$ otherwise $\lambda=0$. This means, in view of Fig.\ref{fig:SplitProductClosedLoops}(b), that the driver is the low-number molecule. If the initial molecule number is equal then the error is  $\lambda$-independent and so we used an equal drive $\lambda =0.5$. For the complex formation equilibrium process the initial order $q_{A0}\lessgtr q_{B0}$ is preserved during time evolution so that either $A$ or $B$ is the driver, not both. When a network is built on many interconnected complex formation processes molecules do not stay in a fixed order at all times. For these networks the $\lambda$-parameters need not be equal to $0$ or $1$ and can take intermediate values between $0$ and $1$.

The error calculated for the equilibrium process reflected the time evolution from the initial state to the equilibrium state. For some combinations of  $k_n$ and initial molecule numbers $q_A$ and $q_B$ the transition regime to equilibrium is very short so the error is more reflective of the equilibrium state. At the end of Sec.\ref{sec:Section2} we study the LC method applied to a dynamical system that is out of equilibrium.

\section{\label{sec:Section2} Elementary Units}

\begin{figure}[h]
\includegraphics[scale=0.3]{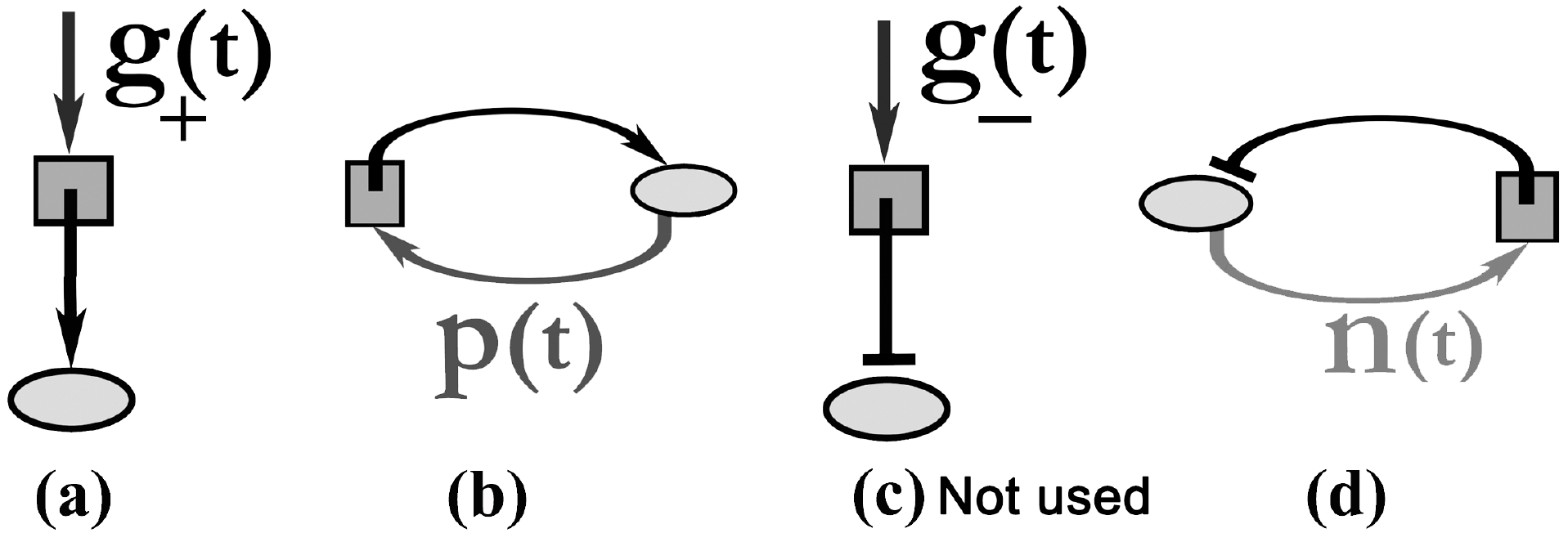}
\caption{\label{fig:ComponentLibrary1} Building blocks with one molecule, $q_A$. The degradation represented in (c) will not be used as an elementary process whereas the other three processes will be used. The transitions for (a) and (b) increase the number of molecules from $q_A$ to $q_A+1$ whereas for (c) and (d) decrease it to $q_A-1$. The terms in the master equation that correspond to the one-molecule processes for (a), (b) and (d) are $g_{+}(t) (z-1) F$, $p(t) (z-1) z \partial_{z} F$ and $n(t) (z^{-1}-1) z \partial_{z} F$ respectively}
\end{figure}

The list of elementary processes contains three more elements besides the complex association ,$A+B\xrightarrow{kp}C$, and the complex dissociation, $C\xrightarrow{kn}A+B$. Fig.\ref{fig:ComponentLibrary1}(a) represents an accumulation process controlled either by the environment or through coupling with another network, both represented by $T_{\epsilon}(q_A,t)= g_{+}(t)$. Another accumulation, Fig.\ref{fig:ComponentLibrary1}(b) with $T_{\epsilon}(q_A,t)= p(t)q_A$, is driven by the molecule itself. The externally controlled degradation from Fig.\ref{fig:ComponentLibrary1}(c) requires a special boundary condition for $P(q,t)$ because the transition probability $T(q_A,t)=g_{-}(t)$ does not automatically become zero when $q_A=0$. The form of the master equation thus needs to be changed for the special case $q_A=0$, so we will not use Fig.\ref{fig:ComponentLibrary1}(c) as an elementary unit because the boundary condition makes the model hard to solve for large networks.

However, a boundary condition-free externally controlled degradation of a molecule can be achieved through the complex formation process Fig.\ref{fig:SplitProductClosedLoops}(a). Consider that $B$ represents the entrance port through which the environment controls the degradation of $A$. The external control may be delivered either through a time-variable coupling $k(t)$ or through the time-variation of  $F_B(t)$ and $F_{BB}(t)$ modulated by the environment or another biocircuit that couples into $B$. For this application the complex $C$ is of no importance. The last elementary process, also free of boundary conditions, is the auto-degradation Fig.\ref{fig:ComponentLibrary1}(d) with $T(q_A,t)=n(t) q_A$.

\begin{figure}[h]
\includegraphics[scale=0.50]{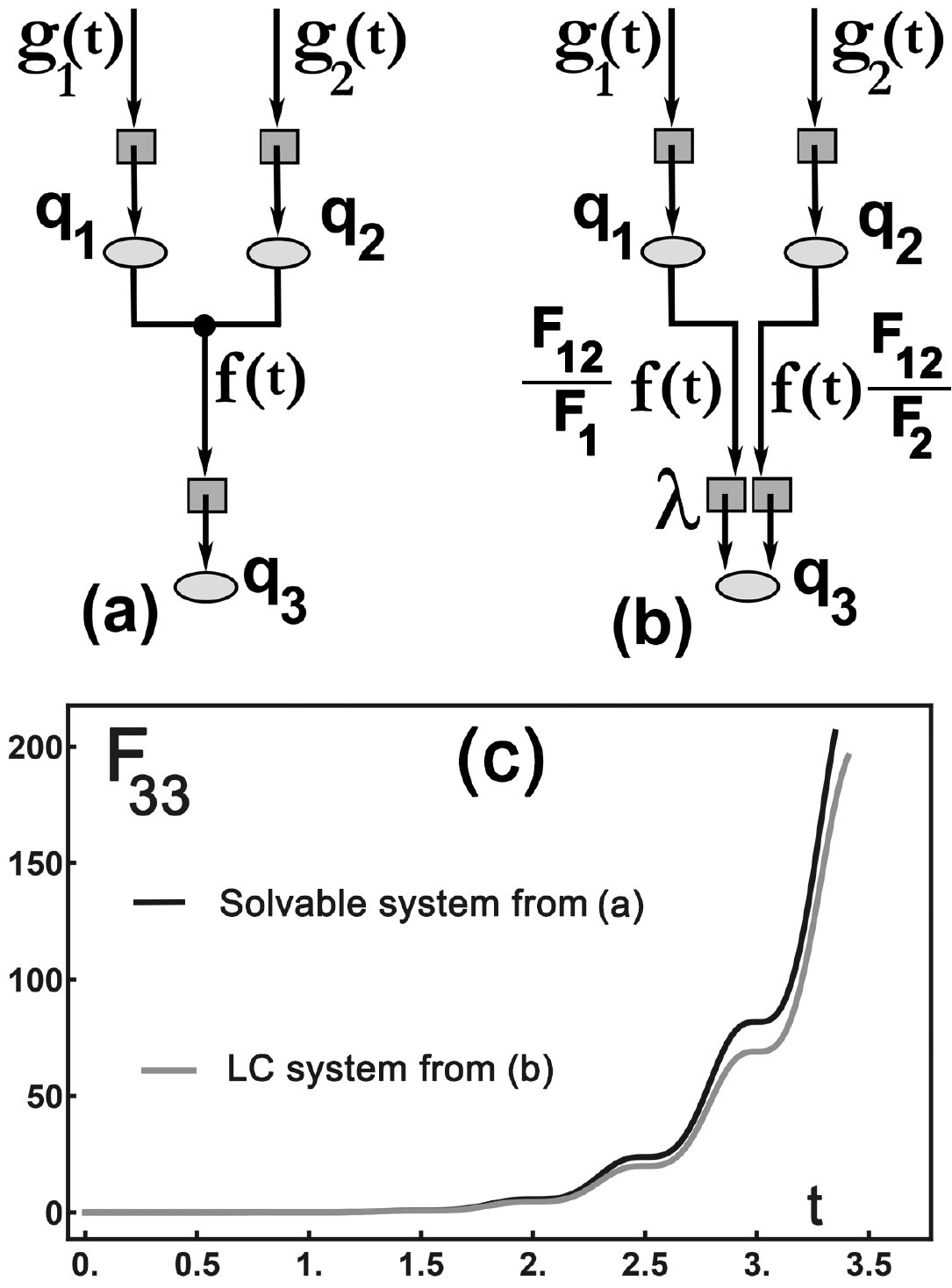}
\caption{\label{fig:FigGenProd.pdf} (a) A $4^{\text{th}}$-order-moment completely solvable network. (b) The product node is split and  the driving parameter $\lambda$ is associated with $q_1$ and $1-\lambda$ with $q_2$. (c) The LC-mean value $F_{33}$ coincides within the exact solution with a mean error of $1.2 \times 10^{-1}$ over the time interval $[0,3.5]$.  The driving molecule is $q_1$ which starts at $t=0$ from zero. The other parameters are, $q_2=1,\; q_3=0$ at $t=0$,\; $g_1(t)=t,\;  g_2(t)=1$, and $f(t)=\sin(2\pi t)^2$.}
\end{figure}

An immediate application of the generators is to build a system that does not settle at an equilibrium state,  Fig.\ref{fig:FigGenProd.pdf}(a). The generators $g_1(t)$ and $g_2(t)$ continually increase the number of molecules $q_1$ and $q_2$ which, in turn, produce more complex $q_3$.  The complex formation transition probability per unit time, $T(q,t)=f(t) q_1 q_2$, is time dependent through $f(t)$ in addition to its dependance on the stochastic time-dependent variables $q_1$ and $q_2$. The network from  Fig.\ref{fig:FigGenProd.pdf}(a) is an example for which the moment equations close in the fourth order and there is no need for a stochastic simulation to estimate them \cite{Narvik}. Because it is solvable, this gives us a chance to study the accuracy of its LC approximation, Fig.\ref{fig:FigGenProd.pdf}(b). For this example the generators $g_1(t)$ and $g_2(t)$ depend on time and the system can be driven into a variety of trajectories. In Fig.\ref{fig:FigGenProd.pdf}(c) we choose $f(t)=\sin(2 \pi t)^2$ to model a coupling on $q_3$ that oscillates between a maximum strength and zero. The error for the mean value $F_3(t)$ is on the order of $10^{-7}$. Maximum errors on the order of $10^{-1}$ appear for the second order moments Fig.\ref{fig:FigGenProd.pdf}(c).  In general, the maximum error cover a range from   $10^{-3}$ to $10^{-1}$  \cite{suppFourthOrder}.

Other, more complicated processes are expressed in terms of the elementary ones. For example a simultaneous collision of 3 molecules would produce a transition probability proportional with $k q_1q_2q_3$ that can be split, but the LC master equation will involve third order moments. Instead, the triple product $k q_1q_2q_3$ can be expressed as a more likely process of sequential collisions in which two molecules collide to form a complex and then this complex collides with a third molecule. This approach will close the LC equations at second order.

Common types of transition probabilities are built out of rational functions. An example is $(1+(q_A)^4)^{-1}$ which represents a gate that closes for large $q_A$. This is not an elementary process because rational-function transition probabilities describe the phenomenological behavior of sub-networks built on elementary reactions. One of these network responsible for ultrasensitivity will be studied in Sec.\ref{sec:Ultrasensitivity}. The advantage of all selected building processes is that the second order moments evolve in time independently of higher order moments, thus the time evolution closes at second order. Because elementary units closely represent biochemical processes, the selection of the network's topology becomes intuitive.

To exemplify the method described above, in the next two sections we build two networks out of the elementary units and use the LC procedure to split each complex formation control node. The first biocircuit has four nodes and thus it needs four splittings. Each splitting introduces a $\lambda$ parameter. The second one requires ten splittings. The reason for choosing these specific examples is explained below.

\section{\label{sec:level3} Networks with multiple equilibrium states}

The first biocircuit is a bistable network. In multistable regulatory networks noise elicits a phenotypical binary response by driving transitions between distinct locally stable states. The transition can adapt the organism to a change in environment, switching back once the change elapsed \cite{NatureBystable,WongBystable,LobBystable}. Other bistable networks use noise to generate an irreversible cell-fate decision such as hematopoietic cell differentiation. Besides being important as a biological system, we are particularly interested in bistable circuit because it gives the opportunity to reveal the use of the $\lambda$-parameters that appear after splitting the nodes. To this end, consider a bistable network that starts from a given initial state. When analyzed with the deterministic mass-action method it is attracted to one of the two states, but not both.  The same bistable network that starts from the identical initial state as above but now analyzed by stochastic simulations shows trajectories that transition between two distinct locally stable states due to stochastic fluctuations.  The mass-action method cannot reveal this stochastic passage between the equilibrium states, however the LC-method can show the bistability using the $\lambda$-parameters.

\begin{figure}[h]
\includegraphics[scale=0.65]{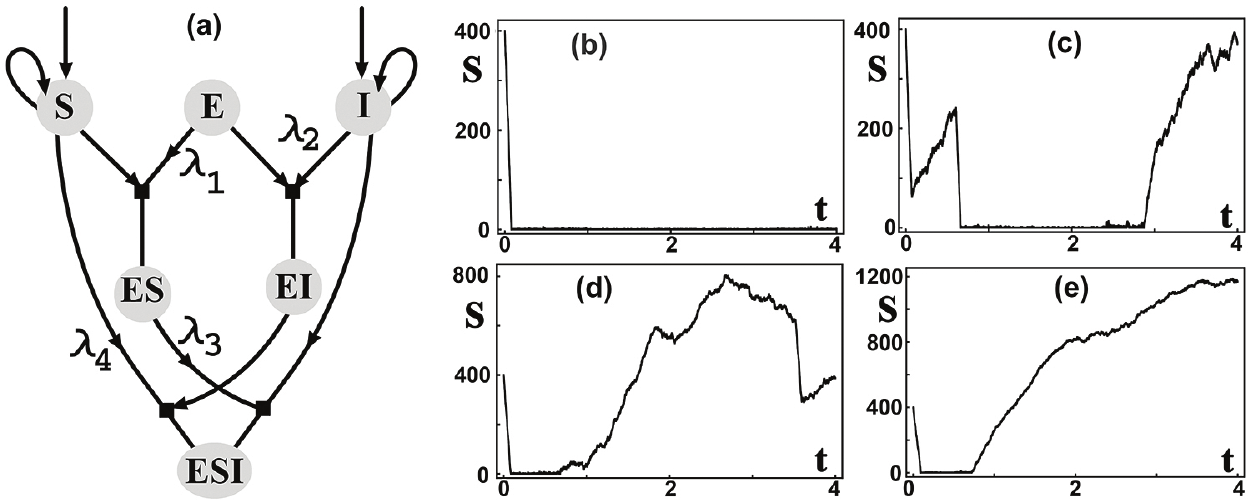}
\caption{\label{fig:figureESI.pdf} (a) The control-action diagram for the bistable system. The driving parameters are associated as follows: $(E,\lambda_1), (I,\lambda_2), (ES,\lambda_3)$ and $(S,\lambda_4)$. (b-e) Four simulated stochastic paths for $S$. All four paths start form $S=400$ molecules but evolve into different trajectories. In (b) the S runs between $0$ and $1$  even if the generator on $S$ is continuously pumping. Contrary to (b), (e) shows a paths that reaches high number of molecules}
\end{figure}

To demonstrate the LC-method applied to bistability we use the bistable system from Fig.\ref{fig:figureESI.pdf} which illustrates the stochastic reactions $E+S\rightleftarrows ES\rightarrow E+P$, $E+I\rightleftarrows EI$ and $ES+I\rightleftarrows ESI\rightleftarrows EI+S$ \cite{CraciunBystable}. We are interested in the behavior of $S$ but not of $P$ and so the reaction $ES\rightarrow E+P$ is not represented in  Fig.\ref{fig:figureESI.pdf}(a). Both $S$ and $I$ are coupled to environment.
The bistability shows itself in the profiles of the $S$-molecule stochastic paths. In Fig.\ref{fig:figureESI.pdf}(b), the molecule $S$ starts from the initial value $F_S(0)=400$ and drops quickly to zero. The environment does not pump enough $S$ molecules into the system to avoid $S$ depletion over the time horizon $(0,4)$. In Fig.\ref{fig:figureESI.pdf} the effect of bistability is visible on three paths. One path starts to rise before $t=1$ and reaches a value of $S=1200$ at $t=4$, Fig.\ref{fig:figureESI.pdf}(e). The paths from Fig.\ref{fig:figureESI.pdf}(c,d) transit between the two states, depletion and high values of $S$. A mass-action deterministic equation using the same numerical parameters and the initial state is unable to reveal the bistability, it only shows the depletion state Fig.\ref{fig:figureESI.pdf}(b).

​

\begin{figure}[h]
\includegraphics[scale=0.65]{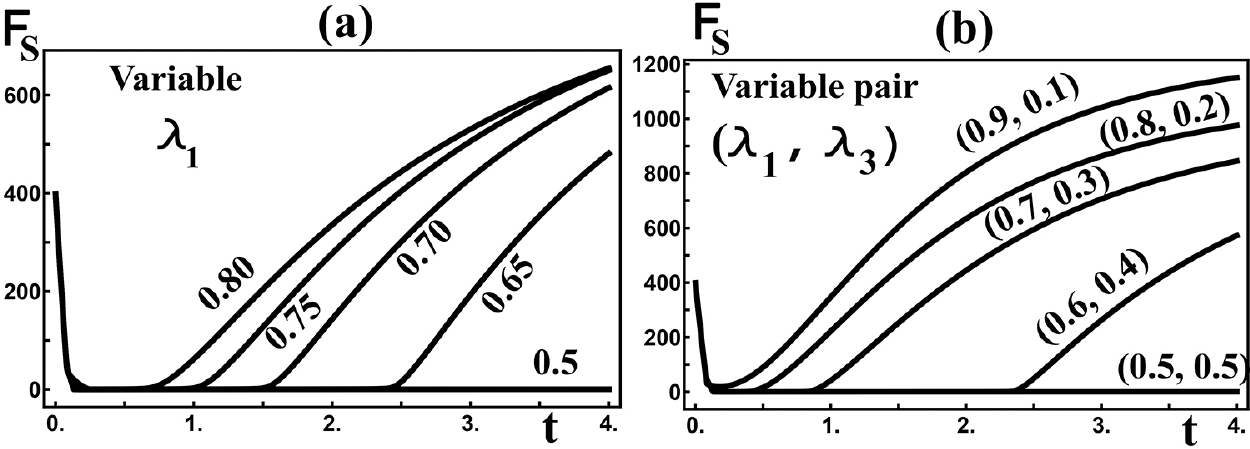}
\caption{\label{fig:PathsESILambdaV1andV2.pdf} In (a) by varying $\lambda_1$ we obtain different transition times for the low to the high state. The rise of $S$ is not much greater than the initial value which mimics Fig.\ref{fig:figureESI.pdf}(c,d). In (b) two parameters are varied $\lambda_1$ and $\lambda_3$. The accumulation of $S$ reaches higher levels than the initial value as in Fig.\ref{fig:figureESI.pdf}(e). All other $\lambda$ are set to $0.5$ except the ones that we varied above.}
\end{figure}

The reason is that the mass-action decouples the mean value equation from the second order moments and the $\lambda$-parameters are lost. Nevertheless, the $\lambda$-parameters show the bistable nature for the mean value of $S$, Fig.\ref{fig:PathsESILambdaV1andV2.pdf}. The average value of $S$ computed from LC ordinary differential equations \cite{Wolfram} shows both the low and the high accumulation states. Different shapes for the mean value of $S$  can be obtained by varying the $\lambda$-parameters, Fig.\ref{fig:PathsESILambdaV1andV2.pdf}. These shapes correspond to the average of different subgroups of stochastic paths that are produced by the bistable phenomenon. The deterministic mass-action result is obtained for $\lambda_1=0.5$ in Fig.\ref{fig:PathsESILambdaV1andV2.pdf}(a).

It may seem that this results from the presence of $\lambda$'s in the mean value equations, however this is not the case. The mean value equations do not depend explicitly on $\lambda$'s. Their dependence on the $\lambda$'s is through the correlation moments that drive the mean values. The $\lambda$'s explicitly drive only the second moments. The necessity of the second moments to reveal the bistability for this network is emphasized by the fact that this example was taken from \cite{CraciunBystable}, where a theorem is provided to help select networks with bistable states. Although the theorem is devised on classical mass-action it distinguishes between some networks that can support bistable behavior and others that cannot. However, for the example from Fig.\ref{fig:figureESI.pdf}(a) the theorem cannot say if it is bistable or not. On the other hand, the LC-method is capable of showing the bistability of  Fig.\ref{fig:figureESI.pdf}(a).

\section{\label{sec:Ultrasensitivity}Ultrasensitivity}

\begin{figure}[h]
\includegraphics[scale=0.9]{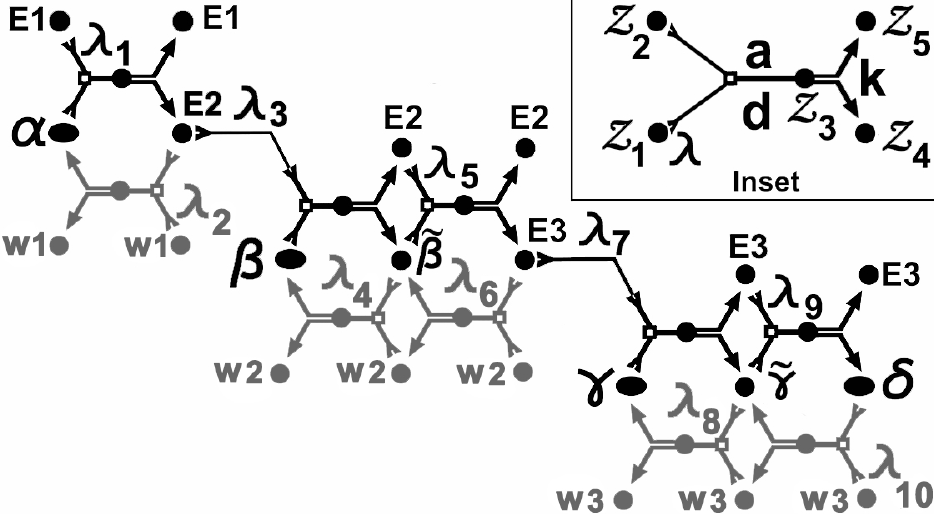}
\caption{\label{fig:FigureMAPKV11.pdf}\kern-1em The driving parameters are associated as follows: $(E1,\lambda_1)$, $(E3,\lambda_3)$, $(E2,\lambda_5)$, $(E3,\lambda_7)$, $(E3,\lambda_9)$ for the downstream cascade and $(w3,\lambda_{10})$, $(\widetilde{\gamma},\lambda_8)$, $(E3,\lambda_6)$, $(\widetilde{\beta},\lambda_4)$, $(E2,\lambda_2)$ for the upstream cascade. The same molecule may drive more than one complex formation. The cascade is assembled by linking the module from the inset, which describes the reaction $z_1+z_2\rightleftarrows z_3\rightarrow z_4+z_5$. Here $z$ stands for the corresponding molecule.}
\end{figure}

An ultrasensitive network delivers a binary (ON-OFF) output which is useful for decision-making processes. The output switches from ON to OFF if the input crosses a threshold value. The ultrasensitive network  acts to filter out small stimuli below the threshold and so understanding its stochastic properties are important for designing switches that avoid accidental  triggering events. In \cite{UltrasensitivityWeiss}  an ultrasensitive synthetic transcriptional cascade was constructed where it was noted that a proper matching of the kinetic rates of the cascade's elements are crucial for a clear separation between the ON and OFF states. The design and the construction of a noise-tolerant ultrasensitive biocircuit was reported in \cite{UltrasensitivityMoon}.
Ultrasensitivity can be achieved by more than one network topology. Here we study one possibility, Fig.\ref{fig:FigureMAPKV11.pdf} based on \cite{FerrellHuang}, for two reasons. First, to test the LC-method on a network that needs hundreds of differential equations for its time evolution. For this example a total of 275 moments are needed, out of which 22 are mean values, and 231 are correlations. Second, given that the number of correlations increases quadratically with the number of molecules, we discuss procedures that project out molecules in order to reduce a network to a simpler one.

The equation for the Inset in Fig.\ref{fig:FigureMAPKV11.pdf} is
\begin{equation}\label{eq:MAPK}
\begin{split}
  \partial_{t}F=\lambda (z_1^{-1}z_2^{-1}z_3-1)\frac{F_{12}(t)}{F_1(t)}a\, z_1\partial_{z_1}F\\
  +(1-\lambda)(z_1^{-1}z_2^{-1}z_3-1)\frac{F_{12}(t)}{F_2(t)}a\, z_2\partial_{z_2}F\\
 +(z_1 z_2 z_3^{-1}-1)d\, z_3\partial_{z_3}F \\ +(z_4 z_5 z_3^{-1}-1)k\, z_3\partial_{z_3}F
\end{split}
\end{equation}

The equation for the entire network is obtained by summing ten terms, each being similar to (\ref{eq:MAPK}). The stochastic time evolution will thus include ten $\lambda-\text{parameters}$.

\begin{figure}[h]
\includegraphics[scale=0.65]{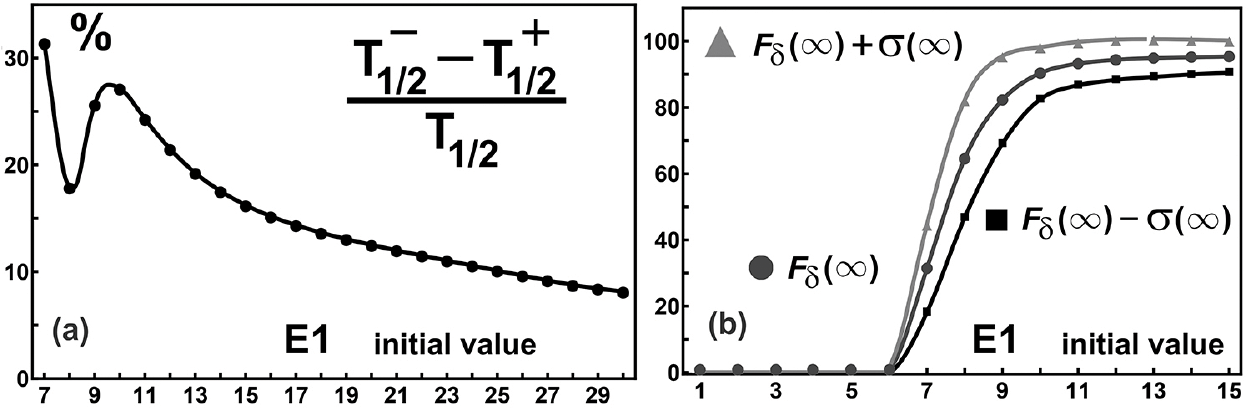}
\caption{\label{fig:CompositeForMAPKSigmoidFlatten.pdf}Response time for the ultrasensitive network.}
\end{figure}

The response time, $T_{1/2}$, of the ultrasensitive switch is one out of many specific time-evolution properties which can be retrieved from the 275 moments. $T_{1/2}$, which depends on each stochastic path, represents the  time for the molecule $\delta$ to reach $1/2$ of its equilibrium level. It plays a central role in network communication. For example, if $\delta$ controls a subsequent pulse-generating network the duration of the generated pulse depends on the controlling $\delta$'s $T_{1/2}$. The pulse may even be absent if the response time is too small and so the range of values for $T_{1/2}$ is relevant to signalling. The range, $T_{1/2}^{-}-T_{1/2}^{+}$, where $T_{1/2}^{\pm}$ represent the response times of  the average evolution of $\delta$ $\pm$ one standard deviation, is computable with the LC method Fig.\ref{fig:CompositeForMAPKSigmoidFlatten.pdf}(a). The maximum relative range in Fig.\ref{fig:CompositeForMAPKSigmoidFlatten.pdf}(a) occurs for the input $E1=7$. Below $E1=7$ the switch is not opened and the response time $T_{1/2}$ is meaningless. At $E1=7$ the switch is just about to open as can be seen in Fig.\ref{fig:CompositeForMAPKSigmoidFlatten.pdf}(b) which illustrates the sigmoidal dependance of the equilibrium output mean value, $F_{\delta}(\infty)$, in terms of the initial input $F_{E1}(0)$. Flanking the mean value response are the equilibrium responses $F_{\delta}(\infty)\pm\sigma(\infty)$ which highlight the stochastic nature of the sigmoidal response. A local maximum in Fig.\ref{fig:CompositeForMAPKSigmoidFlatten.pdf}(a) appears for $E1$ between $9$ and $11$ when the ultrasensitive system is just about to be fully opened.

\section{Reducing a network by splitting and projection}

\begin{figure}[h]
\includegraphics[scale=0.65]{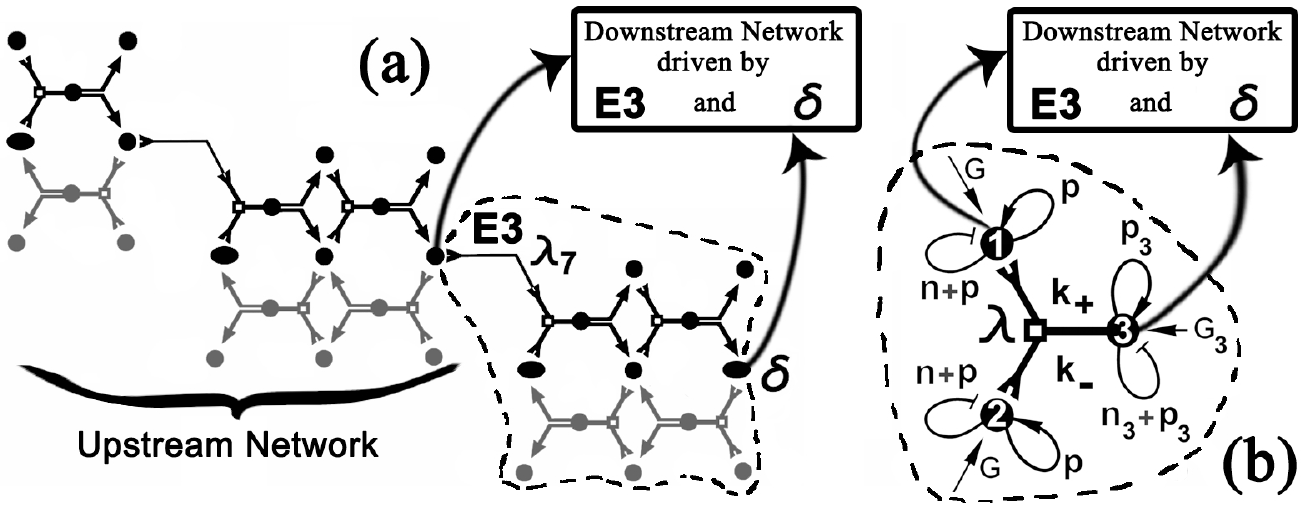}
\caption{\label{fig:FigureMAPKV16.pdf}(a) The dotted line encircles a subnetwork that drives the downstream network and is driven by the upstream subnetwork. (b) Two-molecules equivalent system. The input molecules, $1$ and $2$, will have identical time evolution. They start with the same initial conditions and are driven by the same $G$, $n$ and $p$ external actions. The output molecule, $3$, is controlled by $G_3$, $n_3$ and $p_3$. Here $\lambda =0.5$ because the molecules $1$ and $2$ have identical evolution.}
\end{figure}

Recall from Sec. II that the LC method relies on the equivalence of Fig.\ref{fig:Equilibrium}(a) and Fig.\ref{fig:SplitProductClosedLoops}(a) up to second order. The broad idea of splitting biocircuits to create equivalent systems can be applied further to large networks like the one seen in Sec.\ref{sec:Ultrasensitivity}. A subnetwork can be disconnected from a larger network and subsequently simplified to create a smaller, simpler equivalent system. We will use the encircled subnetwork in Fig.\ref{fig:FigureMAPKV16.pdf}(a) as an example.

Consider that two molecules from the ultrasensitive cascade, $E3$ and $\delta$, drive a downstream network which does not influence the dynamics of the ultrasensitive cascade through any feedback Fig.\ref{fig:FigureMAPKV16.pdf}(a). Moreover, the subnetwork encircled by the dotted line in Fig.\ref{fig:FigureMAPKV16.pdf}(a) only influences the downstream network without influencing the upstream network. Ideally, the information that flows from $E3$ and $\delta$ into the downstream network would be confined exclusively to the five moments $M_5=(F_{E3}, F_{\delta}, F_{E3\delta}, F_{E3E3}, F_{\delta\delta})$. If that would happen, the time evolution of the downstream network could be easily decoupled from the ultrasensitive network. However, the differential equations that describe the evolution of the downstream network contain moments of the ultrasensitive network other than those aforementioned. Many correlations internal to the upstream network will couple through $E3$ and $\delta$ into the downstream network. Thus, we will set up a simplified model for the encircled subnetwork and then fit this model to the data given by  known $M_5$. The moments $M_5$ are known from solving the ultrasensitive network from Sec.\ref{sec:Ultrasensitivity}.

To simplify the input into the downstream network we disconnect the encircled subnetwork and reduce it to the simpler topology from Fig.\ref{fig:FigureMAPKV16.pdf}(b). The simpler topology is not unique. There are many ways to make the reduction. For the specific reduction presented in Fig.\ref{fig:FigureMAPKV16.pdf}(b) our reasoning is as follows. If only one molecule is used in the reduced system it would produce two variables $F_1$ and $F_{11}$ and be driven by a maximum of three elementary units Fig.\ref{fig:ComponentLibrary1}. This topology is too small to fit this model to the five moments of the driving molecules $M_5$. If we were to use two distinct molecules we get five moments and six elementary units, three per molecule, which is enough to fit the model to the five moments. Using three or more molecules would be even easier to accommodate five moments, but we start to lose the simplicity of the equivalent network.

In Fig.\ref{fig:FigureMAPKV16.pdf}(b) we settled on using two molecules $1$ and $2$, which are identical, and the complex, $3$, which is a dimer formed from $1$ and $2$. Usually the complex formation has three distinct molecules, however we have set $1$ and $2$ as identical to ensure we have the least number of different types of molecules possible while having enough elementary units for optimization. We chose this topology because complex formation is a repeated pattern in the ultrasensitive network. The complex formation introduces two unknowns, the association and dissociation coefficients $k_{+}$ and $k_{-}$. The autodegradation function here is written as $n(t)+p(t)$ instead of $n(t)$ because together with the autoaccumulation $p(t)$ it drives the time evolution as  $(z-1) z p(t) \partial_z F+(z^{-1}-1) z p(t) \partial_z F+(z^{-1}-1) z n(t) \partial_z F$. Thus, $p(t)$ acts as a diffusion process $(z-1) z p(t) \partial_z F+(z^{-1}-1) z p(t) \partial_z F$. The advantage of using a diffusion plus a negative autoregulation instead of a positive and a negative autoregulation is that the diffusion $p(t)$ does not affect the mean value, it changes only the standard deviation. The same logic is used for $p_3(t)$. In this way the diffusion terms $p(t), p_3(t)$ act mainly either around the initial time or when the system leaves the transitory regime and enters into the equilibrium state \cite{suppFourthOrder}.
Once the topology is defined the unknowns, $k_{+}$, $k_{-}$, $G(t)$, $p(t)$, $n(t)$, $G_3(t)$, $p_3(t)$ and $n_3(t)$, shown in Fig.\ref{fig:FigureMAPKV16.pdf}(b) are found through fitting the model to the given five moments $M_5$. We used Mathematica \cite{Wolfram, suppReducingANetwork} to minimize the error subject to the evolution constraints $F_{E3}(t)=F_1(t)$, $F_{\delta}(t)=F_3(t)$, $F_{E3\delta}(t)=F_{13}(t)$, $F_{E3E3}(t)=F_{11}(t)$ and $F_{\delta\delta}(t)=F_{33}(t)$. Molecule $2$ is not part of the minimization constraints because it is identical with molecule $1$.
With this strategy, we project out $7$ of $9$ molecules that are in-between $(E3,\lambda_7)$ and $\delta$ in the original ultrasensitive network.

\section{Conclusions}

We have shown that the mathematics and the diagrams of biocircuits are in fact interdependent and can be used to give an accessible method that produces quantitative results from qualitative pictures. By modelling larger interactions as combinations of the elementary units, networks that span to hundreds of interactions can be built. Importantly, the results maintain their stochastic nature as all of the equations come from the Pauli master equation. This saves the oftentimes huge computational expense of running a stochastic simulation which becomes impractical for large systems. The master equation also plays into the ease of the method because for each action in the diagrams, there is corresponding term in the equation.

The terms in the master equation from the split product gave rise to the $\lambda$-parameters. The $\lambda$-parameters revealed that the low molecule species is the driver in product interactions. They also allowed us to investigate different paths in bistability. Selecting different $\lambda$'s allowed the selection of different paths in the bistable process.

Different directions lay ahead for future studies. Instead of taking the initial limit in the updating process of $\Delta t\rightarrow 0$ we can keep $\Delta t$ finite and let the updating process run at discrete times. Moreover, the time intervals $\Delta t$ for each update do not need to be equally spaced and, even more they may be drawn from a probability distribution. In this way some elementary units or subnetworks will be updated more often than others. This type of approach is similar with part of the Gillespie algorithm for which the time between reaction is stochastic. In this case the LC-method becomes a hybrid, keeping the differential equations for the reactions but the time updating process needs stochastic simulations.



\bibliography{apssampLT15380}




\clearpage
\widetext
\begin{center}
\textbf{\large Supplemental Materials: Splitting Nodes and Linking Channels: A Method for Assembling Biocircuits from Stochastic Elementary Unit}
\end{center}

\setcounter{equation}{0}
\setcounter{figure}{0}
\setcounter{table}{0}
\setcounter{page}{1}
\setcounter{section}{0}
\makeatletter
\renewcommand{\theequation}{S\arabic{equation}}
\renewcommand{\thefigure}{S\arabic{figure}}
\renewcommand{\bibnumfmt}[1]{[#1]}
\renewcommand{\citenumfont}[1]{#1}

\section{\label{SEquilibriumReaction}Equilibrium Reaction}

\begin{figure}[h]
\includegraphics[max size={\textwidth}{\textheight}]{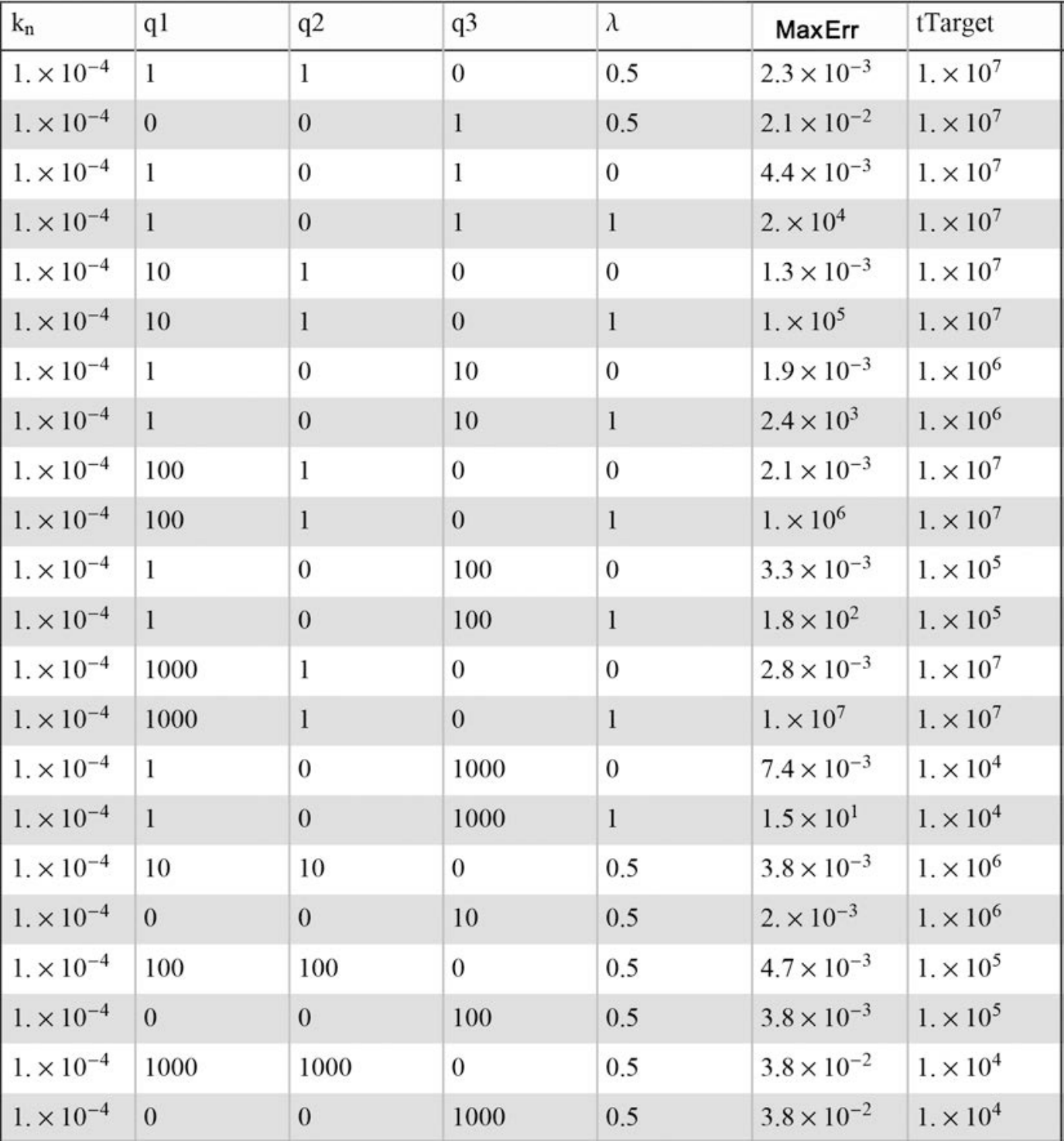}
\caption{\label{fig:TableEquilibriumKnMinus4.pdf}Table for $k_n=10^{-4}$}
\end{figure}

\begin{figure}[h]
\includegraphics[max size={\textwidth}{\textheight}]{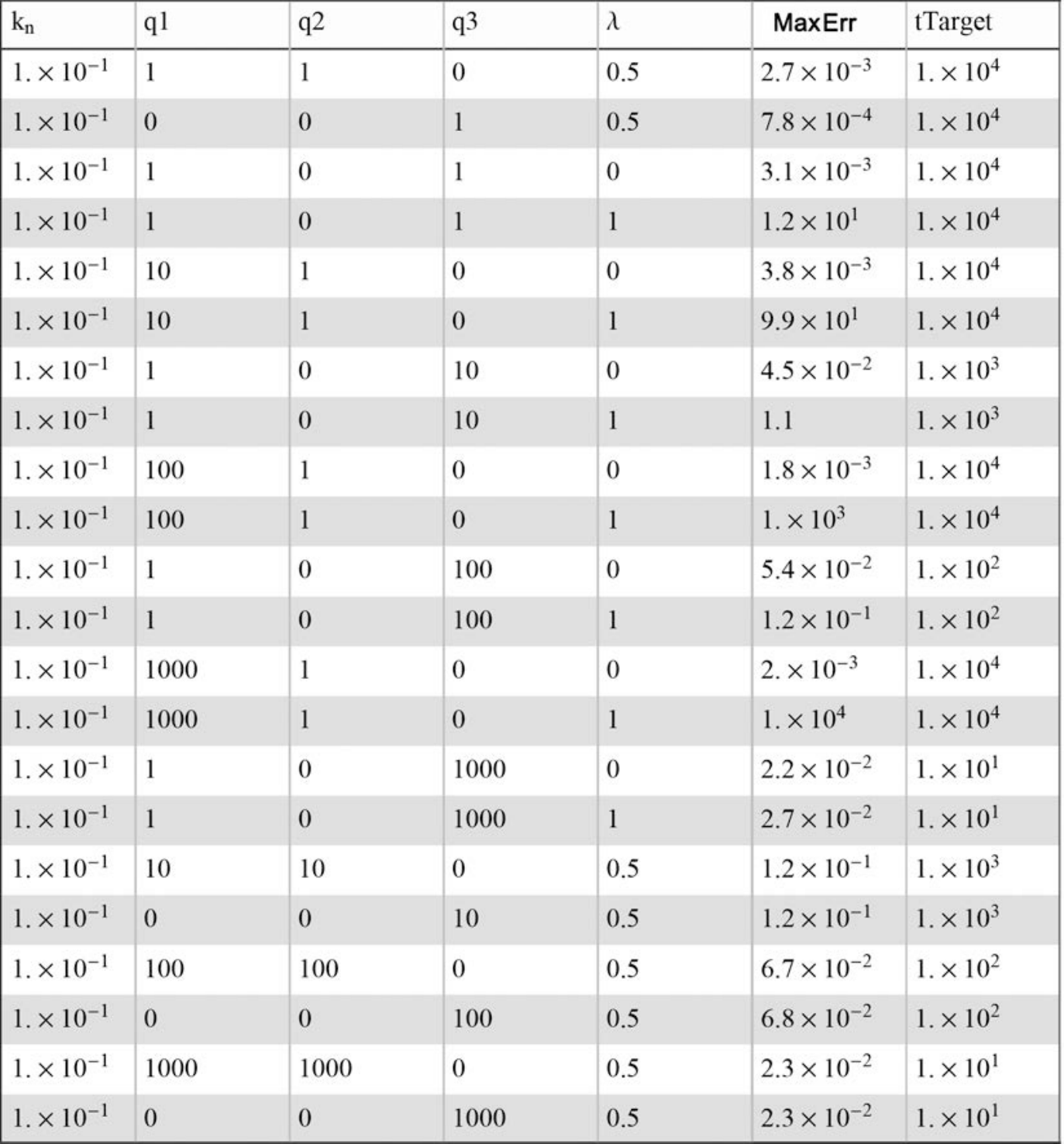}
\caption{\label{fig:TableEquilibriumKnMinus1.pdf}Table for $k_n=10^{-1}$}
\end{figure}

\begin{figure}[h]
\includegraphics[max size={\textwidth}{\textheight}]{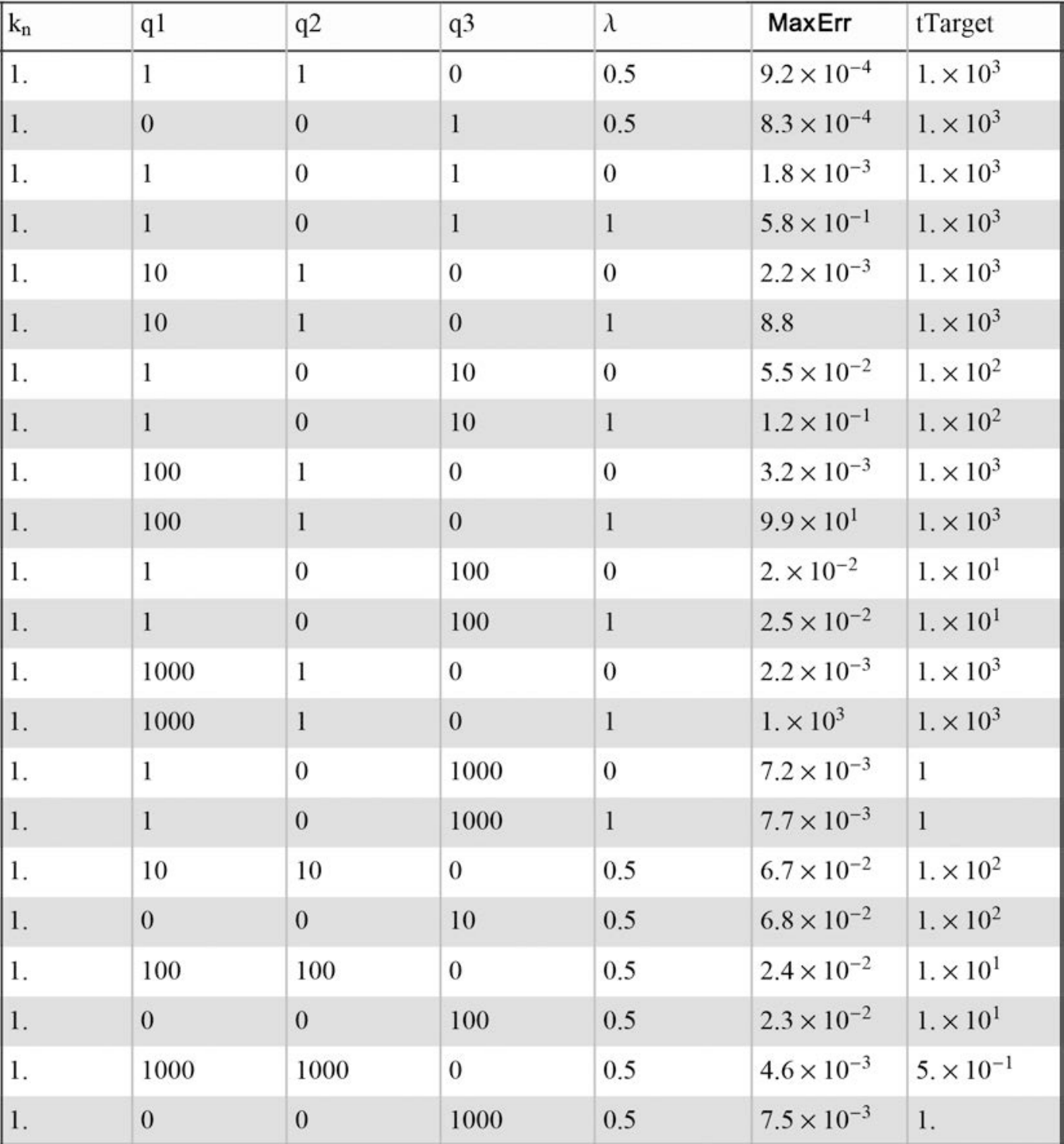}
\caption{\label{fig:TableEquilibriumKn1.pdf}Table for $k_n=1$}
\end{figure}

\begin{figure}[h]
\includegraphics[max size={\textwidth}{\textheight}]{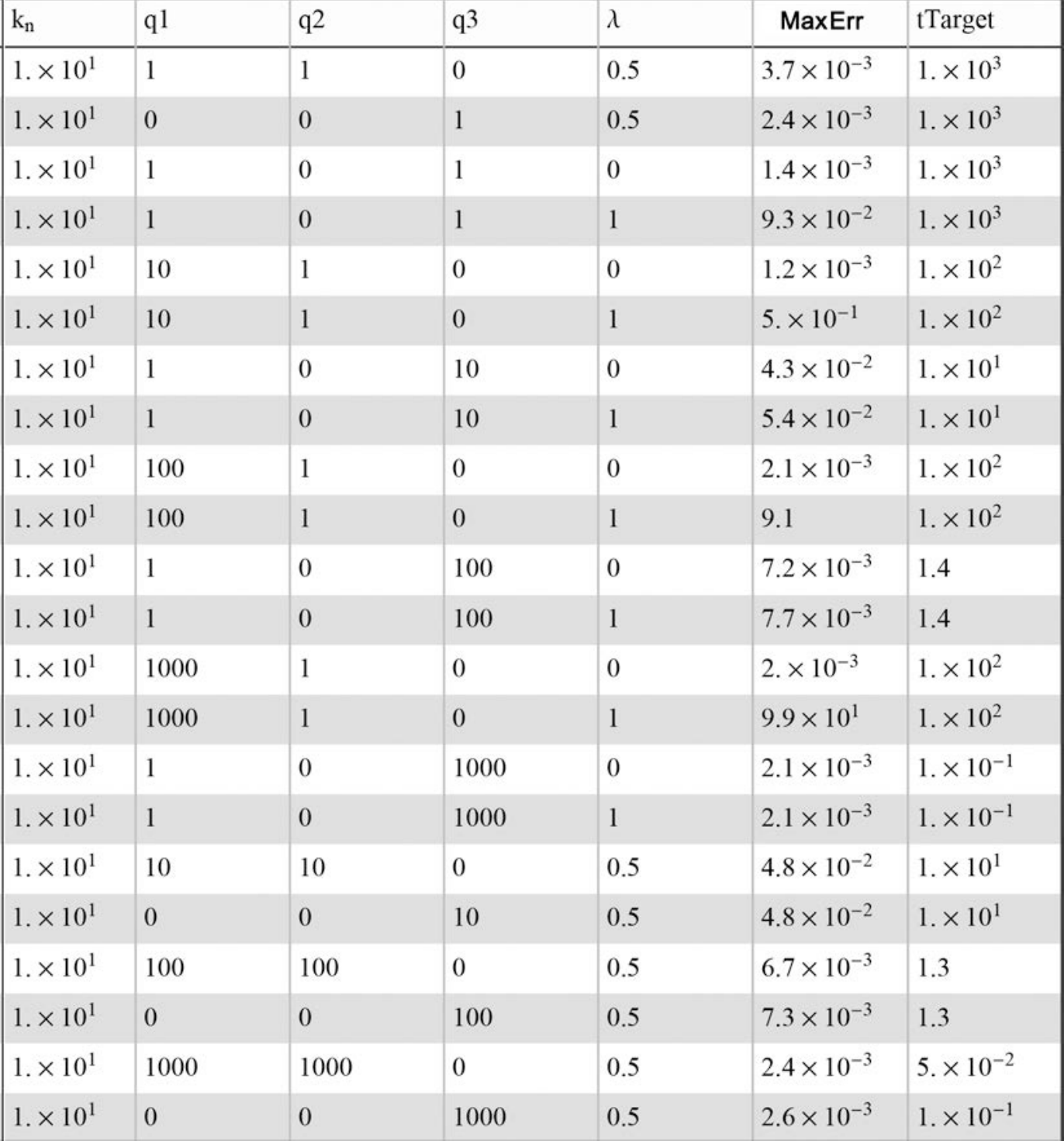}
\caption{\label{fig:TableEquilibriumKnPos1.pdf}Table for $k_n=10$}
\end{figure}

\begin{figure}[h]
\includegraphics[max size={\textwidth}{\textheight}]{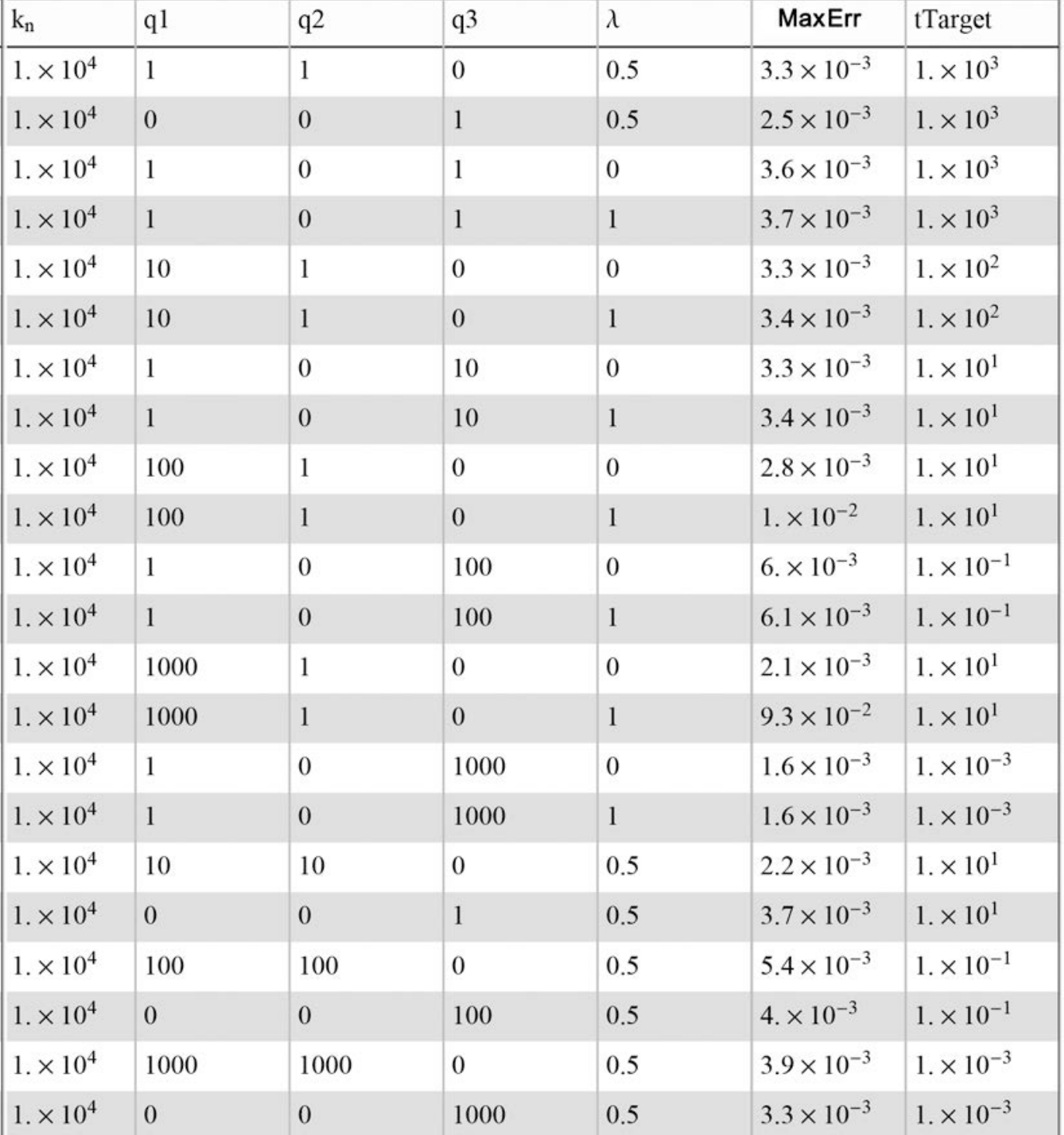}
\caption{\label{fig:TableEquilibriumKnPos4.pdf}Table for $k_n=10^4$}
\end{figure}

In what follows we describe the procedure used to compute the accuracy of the LC-method on complex formation in equilibrium reactions: $A+B\xrightarrow{kp}C$, $C\xrightarrow{kn}A+B$. The final results are tabulated in
Fig.\ref{fig:TableEquilibriumKnMinus4.pdf} to Fig.\ref{fig:TableEquilibriumKnPos4.pdf}. All parameters were varied except $k_p$ which was set to $k_p=1$ so that the time scale in the Master Equation is expressed relative to $k_p$. Changing $k_p$ is equivalent to changing the time unit. To give an example of the tabulated results we used  $k_n=10^{-4}$ and the initial conditions, $q_A=1, q_B=0, q_C=1000$, which gives the initial generating function $F(z_A,z_B,z_C,t=0)=z_A^{1}z_B^{0}z_C^{1000}$. This specific example was chosen because low molecule numbers like $1$ and $0$ are not easy to study using a system of differential equations because their fluctuations are high relative to the mean. We generated $10^3$ paths for the entire process using the Gillespie algorithm. We denote the $k$-th path for molecules $A$, $B$, and $C$ by ${\cal{P}}_{A,k}$, ${\cal{P}}_{B,k}$, and ${\cal{P}}_{C,k}$ where $k=1\dots 10^3$. The time-horizon was set to $tTarget=10^4$ so that the entire transition process, from the initial state to the equilibrium state, was included in our simulation. The jumps for each stochastic realization ${\cal{P}}(\tau^{(k)}_j)_{A,k}$, ${\cal{P}}(\tau^{(k)}_j)_{B,k}$ and ${\cal{P}}(\tau^{(k)}_j)_{A,k}$ appear at $\tau^{(k)}_j$, which are random numbers generated by Gillespie algorithm. The index $j$ starts at $1$ for each $k$ but ends at a random value $j^{(k)}_{max}$ which is dictated by the condition $\tau^{(k)}_{j^{(k)}_{max}+1}>tTarget$. In our simulations $j^{(k)}_{max}$  was about $3500$.

\begin{figure}[h]
\includegraphics[scale=0.3]{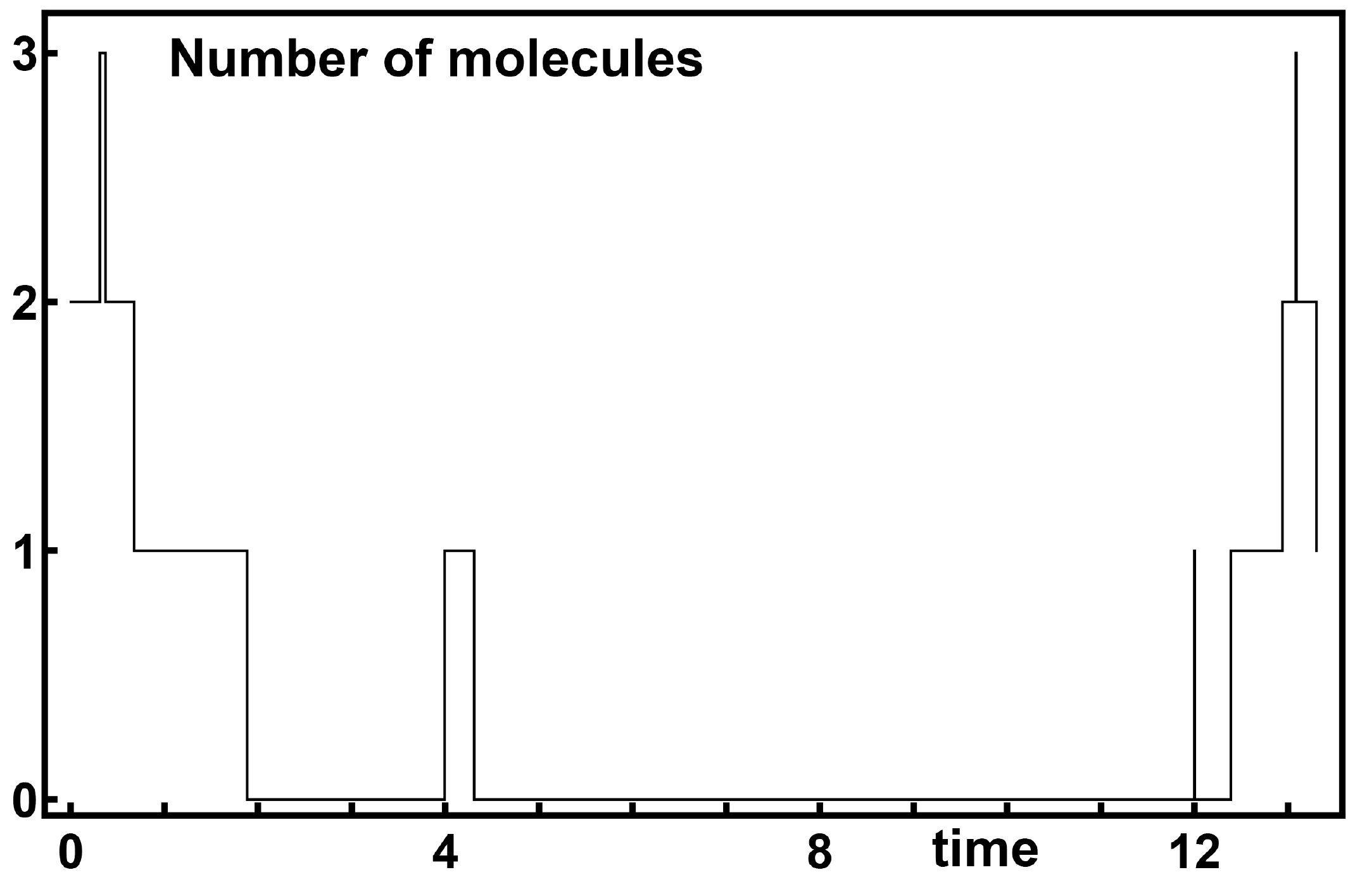}
\caption{\label{fig:SpikyDataV2.pdf}An example of a stochastic process with large variation between different state's durations. The spiky jumps has a duration of $5.0\times 10^{-3}$ whereas the duration of the longest zero state is $7.7$. }
\end{figure}

The LC-generating function produces a system of ordinary differential equations for $F_A(t), F_B(t), F_C(t), F_{AA}(t), F_{BB}(t), F_{CC}(t), F_{AB}(t)$ and $F_{BC}(t)$ which were numerically solved with Mathematica on the time interval $[0,tTarget]$. The LC-results were obtained much faster than the results from the stochastic simulations. To compare the simulations with the LC-results we need to compute time-dependent first and the second order factorial moments out of the $10^3$ paths. We cannot take the path average over $k$ for a fixed time $\tau^{(k)}_j$ because these times are stochastic and thus are not the same for all paths. One way to obtain the moments from the simulated data is to interpolate each path and obtain a function defined for each time in the interval $[0,tTarget]$. Then sample these interpolations at a specified time sequence, say $t_s = s \frac{tTarget}{100}$ with $s=0\dots 100$. The path average over $k$ is simplified because the time sequence $t_s$ is common for all paths $k=1\dots 10^3$. This approach works if the molecule number is large and if the difference between adjacent times $\tau^{(k)}_{j-1}-\tau^{(k)}_j$ does not vary wildly. For small molecule numbers a molecule may jump only between states $q=0$ and $q=1$, a linear or other smooth interpolation will introduce artifacts. The result depends on the location of the sampled times $t_s$. As a consequence we used a zero-order interpolation that represents paths as step functions. For this approach the process is kept discrete, the sampled state at $t_s$ is either 0 or 1, not an artifact intermediate number between $0$ and $1$. Although the interpolation artifact is eliminated there is another problem that needs to be solved. Say the state is $q=0$ between $[\tau^{(k)}_{j-1},\tau^{(k)}_j]$ over a time length of $\tau^{(k)}_{j-1}-\tau^{(k)}_j=7.7$, Fig.\ref{fig:SpikyDataV2.pdf}. Next, the state jumps to $q=1$ for a time length $\tau^{(k)}_{j}-\tau^{(k)}_{j+1}=5.0 \times 10^{-4}$. For such spiky jumps, the probability that $t_s$ will land between $\tau^{(k)}_{j}$ and $\tau^{(k)}_{j+1}$ is very small and so the value $q=1$ is not sampled. If the process is such that the state $q=1$ is short lived for the entire process, then we would get the erroneous result that the average value is zero. These problems are solved if we use a zero-order interpolation and take a time average over a subinterval of $[0,tTarget]$. The time average will capture the short living states and so states like $q=1$ over $\tau^{(k)}_{j}-\tau^{(k)}_{j+1}=5.0\times 10^{-4}$ are not lost.

To compute the time average, the interval $[0,tTarget]$ was divided in $10$ subintervals, $[t_i,t_{i+1}]$, with $i=0\dots 9$. The time-average for each path ${\cal{P}}_k$ over each of the 10 time intervals was computed, ${<\cal{P}}_k>_{[t_i,t_{i+1}]}$. Finally for each $[t_i,t_{i+1}]$, which is common to all paths, we took the average over all paths $k=1\dots 10^3$, that is ${10^{-3}\sum_{k}<\cal{P}}_k>_{[t_i,t_{i+1}]>}$. For the second order moments, like $F_{AB}(t)$, we first multiplied the paths ${\cal{P}}_{A,k}$ and ${\cal{P}}_{B,k}$ which can be done because $A$ and $B$ jump at the same time for path $k$. Then we took the time average. The time average over $[t_i,t_{i+1}]$ of the moments obtained from the simulated data are denoted by $F_{A[t_i,t_{i+1}]}^{\text{Gillespie}}$, $F_{B[t_i,t_{i+1}]}^{\text{Gillespie}}$, $F_{C[t_i,t_{i+1}]}^{\text{Gillespie}}$, $F_{AA[t_i,t_{i+1}]}^{\text{Gillespie}}$, $F_{BB[t_i,t_{i+1}]}^{\text{Gillespie}}$, $F_{CC[t_i,t_{i+1}]}^{\text{Gillespie}}$, $F_{AB[t_i,t_{i+1}]}^{\text{Gillespie}}$, $F_{AC[t_i,t_{i+1}]}^{\text{Gillespie}}$ and $F_{BC[t_i,t_{i+1}]}^{\text{Gillespie}}$.

Next the ordinary differential equations from the LC-model were solved with Mathematica.

\begin{eqnarray*}
  \frac{dF_A}{dt}&=&-k_p F_{AB}+k_n F_C \\
 \frac{dF_B}{dt}&=&-k_p F_{AB}+k_n F_C \\
 \frac{dF_C}{dt}&=&k_p F_{AB}-k_n F_C \\
  \frac{dF_{AA}}{dt}&=&2 k_p(1-\lambda) F_{AB}-2 k_p\lambda \frac{ F_{AA} F_{AB}}{F_A}-2 k_p(1-\lambda) \frac{ F_{AB}^2}{F_B}+2 k_n F_{AC} \\
  \frac{dF_{BB}}{dt}&=&2 k_p\lambda F_{AB}-2 k_p(1-\lambda) \frac{ F_{BB} F_{AB}}{F_B}-2 k_p \lambda \frac{ F_{AB}^2}{F_A}+2 k_n F_{BC}\\
  \frac{dF_{CC}}{dt}&=&2 k_p \lambda \frac{ F_{AB} F_{AC}}{F_A}+2 k_p (1-\lambda) \frac{ F_{AB} F_{BC}}{F_B}-2 k_n F_{CC}\\
  \frac{dF_{AB}}{dt}&=& - k_p\lambda \frac{ F_{AA} F_{AB}}{F_A}- k_p (1-\lambda) \frac{ F_{BB} F_{AB}}{F_B}- k_p \lambda \frac{ F_{AB}^2}{F_A}- k_p (1-\lambda) \frac{ F_{AB}^2}{F_B}+k_n F_{AC}+k_n F_{BC}+k_n F_C\\
  \frac{dF_{AC}}{dt}&=& -k_p (1-\lambda) F_{AB}+ k_p\lambda \frac{ F_{AA} F_{AB}}{F_A}- k_p\lambda \frac{ F_{AB} F_{AC}}{F_A}- k_p (1-\lambda) \frac{ F_{AB} F_{BC}}{F_B}+ k_p (1-\lambda) \frac{ F_{AB}^2}{F_B}- k_n F_{AC}+k_n F_{CC}\\
  \frac{dF_{BC}}{dt}&=& -k_p \lambda F_{AB}- k_p\lambda \frac{ F_{AB} F_{AC}}{F_A}+ k_p(1-\lambda) \frac{ F_{AB} F_{BB}}{F_B}- k_p (1-\lambda) \frac{ F_{AB} F_{BC}}{F_B}+ k_p \lambda \frac{ F_{AB}^2}{F_A}- k_n F_{BC}+k_n F_{CC}
\end{eqnarray*}

The initial conditions are generated, as above, from $F(z_A,z_B,z_C,t=0)=z_A^{1}z_B^{0}z_C^{1000}$. This will give $F_B(t=0)=0$ which cannot be used because the LC equations contain $F_B(t=0)$ as a denominator at $t=0$. To avoid division by zero we took $F_B(0)=10^{-4}$. In general, for zero-molecule initial value, we used a small value for the LC initial conditions. For $F_{BB}(t=0)=q_B (q_B-1)|_{q_B->10^{-4}}$ we used $F_{BB}(t=0)=\text{Abs}(q_B (q_B-1))|_{q_B->10^{-4}}$ to avoid negative numbers given that $F_{BB}$ should be either zero or a positive number. The LC-system of differential equations were numerically solved for $\lambda\in\{0,0.1,0.2,\dots,1\}$.

The time averages of the LC moments over $[t_i,t_{i+1}]$ were computed. These time average values are denoted by $F_{A[t_i,t_{i+1}]}^{\text{LC}}$, $F_{B[t_i,t_{i+1}]}^{\text{LC}}$, $F_{C[t_i,t_{i+1}]}^{\text{LC}}$, $F_{AA[t_i,t_{i+1}]}^{\text{LC}}$, $F_{AB[t_i,t_{i+1}]}^{\text{LC}}$, and so on.

The error for the first moment of molecule $A$, corresponding to the time interval $[t_i,t_{i+1}]$ is computed as

\begin{equation}\label{Error1}
  \text{Error}_A[t_i,t_{i+1}]=\frac{\text{Abs}(F_{A[t_i,t_{i+1}]}^{\text{LC}}-F_{A[t_i,t_{i+1}]}^{\text{Gillespie}})}{F_{A[t_i,t_{i+1}]}^{\text{Gillespie}}}
\end{equation}

if $F_{A[t_i,t_{i+1}]}^{\text{Gillespie}}\neq 0$ and

\begin{equation}\label{Error2}
  \text{Error}_A[t_i,t_{i+1}]=\text{Abs}(F_{A[t_i,t_{i+1}]}^{\text{LC}}-F_{A[t_i,t_{i+1}]}^{\text{Gillespie}})
\end{equation}

for $F_{A[t_i,t_{i+1}]}^{\text{Gillespie}}= 0$.

These same formulas were used for all moments and all 10 time intervals. To get an overall view for the error for the entire process, we computed the mean value over the time intervals of the $\log_{10}\text{Error}$ for each moment and then took the maximum  value over the moments:

\begin{equation}\label{MaxErr}
  \log_{10}(\text{MaxErr})=\max\limits_{\text{over all 9 moments}}(\frac{1}{10}\sum_{i=0}^{9} (\log_{10}(\text{Error}_{\text{moment}}[t_i,t_{i+1}]))
\end{equation}

\begin{figure}[h]
\includegraphics[scale=0.7]{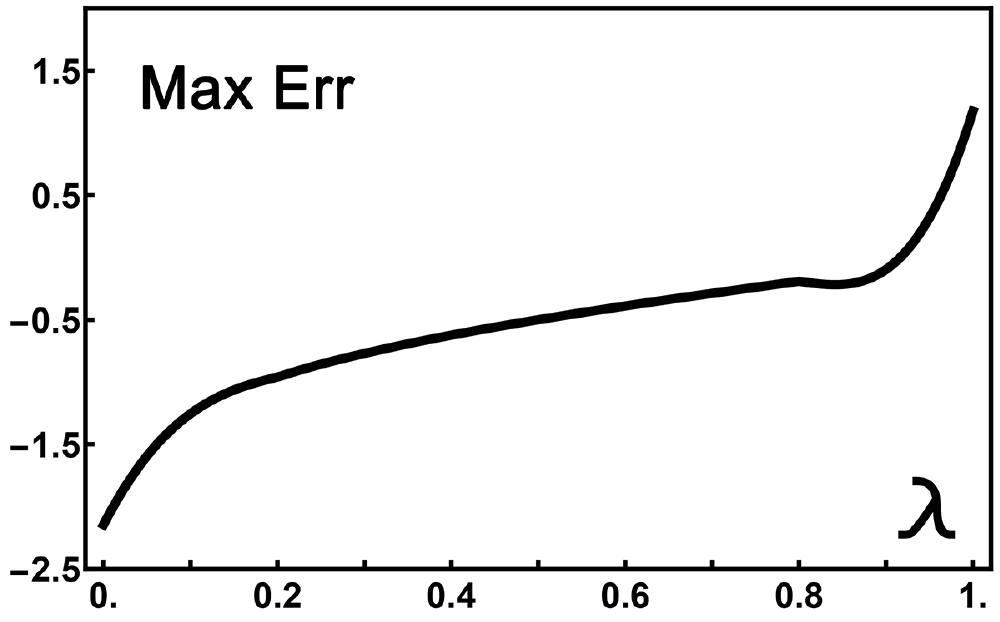}
\caption{\label{fig:fMax.pdf}The maximum error for $k_p=1$,$k_n=10^{-4}$, $q_A=1$,$q_B=0$,and $q_C=1000$. The time horizon is $T=10^4$.}
\end{figure}

MaxErr depends on $\lambda$ Fig.\ref{fig:fMax.pdf}. We noticed that the overall maximum error is lowest for $\lambda=0$. The value $\lambda =0$  eliminates the action of $q_A$ in Fig.\ref{fig:SplitProductClosedLoops}(b) and leaves only the lowest molecule number B, $q_B=0<q_A=1$ in Fig.\ref{fig:fMax.pdf}, as the driving molecule in Fig.\ref{fig:SplitProductClosedLoops}(a). We found that the trend for all cases we simulated and tabulated in Fig. \ref{fig:TableEquilibriumKnMinus4.pdf} to Fig. \ref{fig:TableEquilibriumKnPos4.pdf} was that the LC term that produces the lowest error corresponds to the lowest initial value molecule number.

\newpage
\section{\label{sec:FourthOrderSolution}Fourth order closed solution for the complex formation process}

Here we discuss the complete solution of the biocircuit from Fig.\ref{fig:FigGenProd.pdf}(a). The product node is not split and so the solution extends up to the fourth order before it closes. The molecules that bind, $1$ and $2$, are connected to the environment through the generators $g_1(t)$ and $g_2(t)$, respectively. The finite system of equations contains 13 equations. The moment $F_{33}(t)$ depends on $F_{123}(t)$ which in turns depends on $F_{1122}(t)$. For simplicity, the time argument $t$ is dropped in for the generators, the function $f(t)$ and the moments.

\begin{enumerate}[label=\textbf{\arabic*.}]
\item $dF_1/dt=g_1$
\item $dF_{11}/dt=2 g_1 F_1$
\item $dF_2/dt=g_2$
\item $dF_{22}/dt=2 g_2 F_2$
\item $dF_{12}/dt=g_2 F_1+g_1 F_2$
\item $dF_3/dt=f F_{12}$
\item $dF_{112}/dt=g_2 F_{11}+2 g_1 F_{12}$
\item $dF_{122}/dt=2 g_2 F_{12}+g_1F_{22}$
\item $dF_{1122}/dt=2 g_2 F_{112}+2 g_1 F_{122}$
\item $dF_{13}/dt=f F_{112}+f F_{12}+g_1F_{3}$
\item $dF_{23}/dt=f F_{12}+f F_{122}+g_2 F_{3}$
\item $dF_{123}/dt=f F_{112}+f F_{1122}+f F_{12}+f F_{122}+g_2 F_{13}+g_1 F_{23}$
\item $dF_{33}/dt=2 f F_{123}$
\end{enumerate}

Integrating the system of equations for the moments, the biocircuit's time-evolution can be casted as an input-output mapping. For example, a simple input-output relation become apparent for the mean values of $q_1$ and $q_2$

\begin{equation}\label{F1}
\begin{split}
  &F_1(t)=F_1^0+\int_0^{t} dt_1g_1(t_1)\\&
  F_2(t)=F_2^0+\int_0^{t} dt_1g_2(t_1).
\end{split}
\end{equation}

Here $F_1^0, F_2^0$ are the initial mean values $F_1(0), F_2(0)$ and are considered as the input variables. The output variables are $F_1(t), F_2(t)$.

The input-output relation for all the other moments can be represented as nested integrals. For example, the time evolution of the mean value for the $q_3$ molecule is

\begin{equation}\label{MeanF3}
\begin{split}
F_3(t)=&
\int_0^{t} dt_3 f(t_3) \int_0^{t_3} dt_2\, g_1(t_2) \int_0^{t_2} dt_1\,g_2(t_1)+
\int_0^{t} dt_3 f(t_3) \int_0^{t_3} dt_2\, g_2(t_2) \int_0^{t_2} dt_1\,g_1(t_1)+
\\& F_1^0 \int_0^{t}dt_3 f(t_3) \int_0^{t_3}dt_2\, g_2(t_2)+F_2^0 \int_0^{t}dt_3 f(t_3) \int_0^{t_3}dt_2\, g_1(t_2)+
\\&F_{12}^0\int_0^{t}dt_3 f(t_3)+F_3^0.
\end{split}
\end{equation}

As the transition probability $T(q,t)=f(t) q_1 q_2$ shows, the product $q_1q_2$  controls $q_3$. To make this product visible in $F_3(t)$ we use

\begin{equation}\label{ProductRule}
\begin{split}
&\int_0^{t_3} dt_2\, g_1(t_2) \int_0^{t_2} dt_1\,g_2(t_1)+\int_0^{t_3} dt_2\, g_2(t_2) \int_0^{t_2} dt_1\,g_1(t_1)=\\&
\left(\int_0^{t_3} dt_2\, g_1(t_2) \right)\left( \int_0^{t_3}dt_2\, g_2(t_2)\right).
\end{split}
\end{equation}

and obtain

\begin{equation}\label{MeanF3}
\begin{split}
F_3(t)=&\int_0^{t} dt_3 f(t_3) \left(\int_0^{t_3} dt_2\, g_1(t_2) \right)\left( \int_0^{t_3}dt_2\, g_2(t_2)\right)+
\\& F_1^0 \int_0^{t}dt_3 f(t_3) \int_0^{t_3}dt_2\, g_2(t_2)+F_2^0 \int_0^{t}dt_3 f(t_3) \int_0^{t_3}dt_2\, g_1(t_2)+
\\&F_{12}^0\int_0^{t_4}dt_3 f(t_3)+F_3^0.
\end{split}
\end{equation}

Dropping the integral sign in a nested integral we arrive at a simple notation for the mean value $F_3(t)$

\begin{equation}\label{MeanF3Simple}
F_3(t)=fg_1g_2+fg_2g_1+fg_2F_1^0+fg_1F_2^0+fF_{12}^0+F_3^0.
\end{equation}

Representing the product rule (\ref{ProductRule}) as $g_1g_2+g_2g_1=(g_1g_2)$ we get

\begin{equation}\label{MeanF3SimpleProduct}
F_3(t)=f (g_1g_2)+fg_2F_1^0+fg_1F_2^0+fF_{12}^0+F_3^0.
\end{equation}

Similar formulas can be obtained for all moments \cite{Narvik}.

\section{\label{SolvableUnsplitSplit} The comparison between the solvable system from Fig.\ref{fig:FigGenProd.pdf}(a) and its LC-version}

The moments $F_3(t),F_{33}(t),F_{13}(t)$ and $F_{23}(t)$ were numerically computed by Mathematica for both the solvable system and its LC version. For the solvable system from Fig.\ref{fig:FigGenProd.pdf}(a), we used  the equations from Sec.\ref{sec:FourthOrderSolution} Supplemental Materials. Then the error was computed for each moment:

\begin{equation}\label{Err}
 \log_{10}( \text{Err}_{moment})=\frac{1}{10^3}\sum_{k=0}^{10^3-1} (\log_{10}(\text{Error}_{\text{moment}}(t_k))
\end{equation}

\begin{figure}[h]
\includegraphics[scale=0.65]{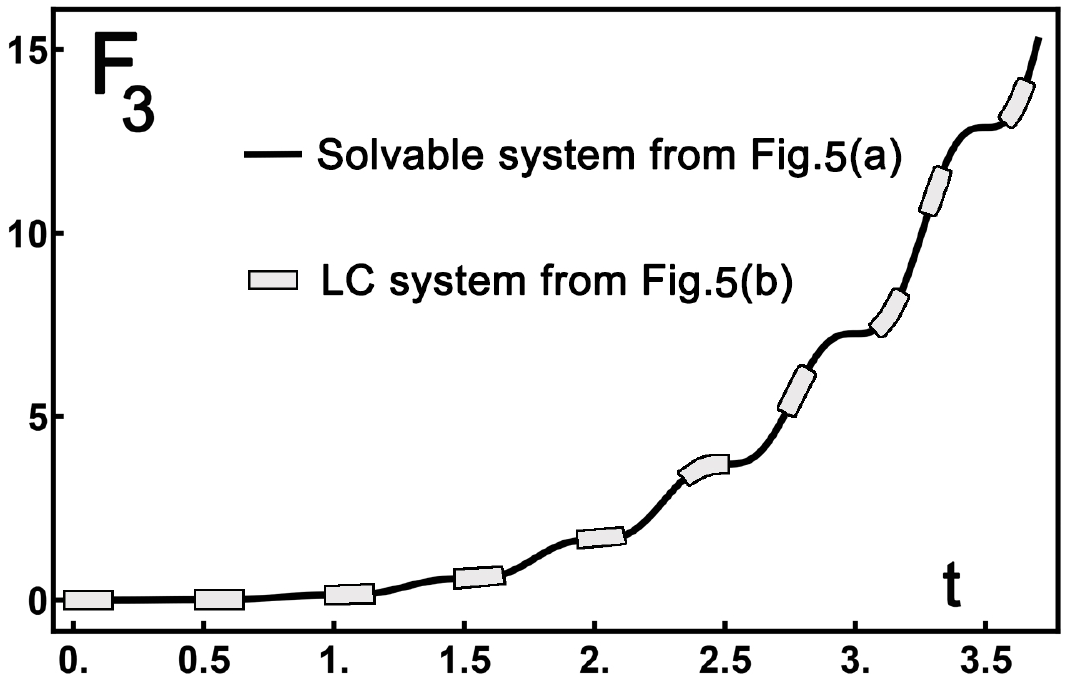}
\caption{\label{fig:FigureUnsplitVersusSplitSinF3V3.pdf} The LC-mean value $F_3$ coincides within the exact solution with a mean error of $2.8 \times 10^{-7}$ over the time interval $[0,3.5]$ }
\end{figure}

\begin{figure}[h]
\includegraphics[scale=0.65]{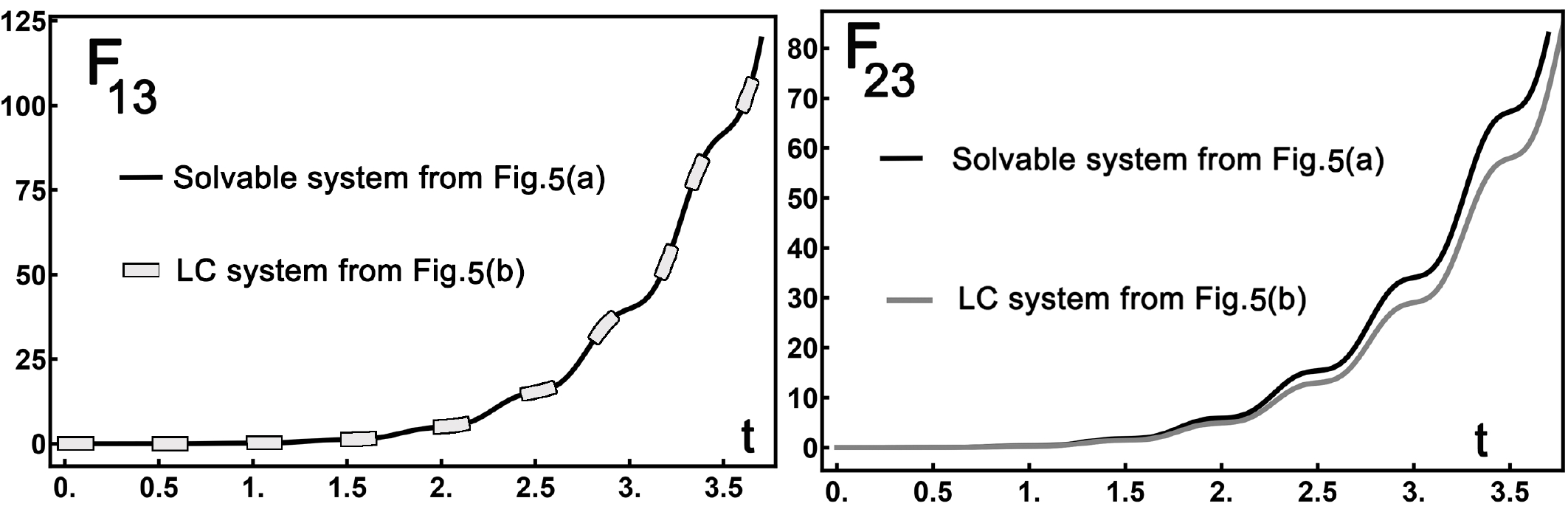}
\caption{\label{fig:FigureUnsplitVersusSplitSinF13F23V3.pdf}  The LC-mean value $F_{13}$ coincides within the exact solution with a mean error of $2.7 \times 10^{-7}$ . The driving molecule is $q_1$, $\lambda=1$, which starts at $t=0$ from zero.  The LC-mean value $F_{23}$ coincides within the exact solution with a mean error of $1.5 \times 10^{-1}$ . The molecule $q_2$  starts at $t=0$ from $1$ and its controlling term is absent from the LC-master equation, $1-\lambda=0$. The error is much smaller for the correlation $F_{13}$ of $q_3$ with the driving molecule $q_1$ then $F_{23}$ between $q_3$ and $q_2$. This situation was observed in many other comparisons between the LC and the exact results }
\end{figure}

Here $\text{Error}_{\text{moment}}(t_k)$ follows the same rule as above, (\ref{Error1}) and (\ref{Error2}), with $F_{\text{moment}}^{\text{Gillespie}}(t)$ being exchanged with $F_{\text{moment}}^{\text{solvable model}}(t)$. Instead of using averages over time intervals $[t_i,t_{i+1}]$ that were required by the stochastic simulation results, here we used numerical values computed at a sequence of time points $t_k=0.1 +k\; (tTarget)/10^3 ,\;k=0\dots 10^3-1$ for both the solvable model Fig.\ref{fig:FigGenProd.pdf}(a) and its LC-approximation Fig.\ref{fig:FigGenProd.pdf}(b). In Figs.\ref{fig:FigureUnsplitVersusSplitSinF3V3.pdf}, and \ref{fig:FigureUnsplitVersusSplitSinF13F23V3.pdf}, $tTarget=3.5$.

We also used constant generators to find the error, by varying $g_1$ and $g_2$ independently  between $0$ and $10$ in steps of $5$. We kept $f(t)=1$. The initial values for $q_1$ and $q_2$ were varied independently choosing the values $0$, $1$, $10$, and $100$. The initial value of $q_3$ was $0$ for all the runs. $\lambda$ was fixed by the driving molecule: $\lambda=1$ if the initial $q_1$ was less then the initial $q_2$. The time horizon was $tTarget=10$. For each of the combinations of the above parameters, we computed $\text{Err}_{moment}$ from (\ref{Err}) for $F_3(t), F_{33}(t), F_{13}(t), F_{23}(t)$. Then we find the maximum value over the 4 moments. This maximum value for all the parameter combinations covered the range from $10^{-3}$ to $10^{-1}$.

\section{The bistable network}

In what follows we describe the LC-method applied on the bistable system from Fig.\ref{fig:figureESI.pdf}. The reactions, the transition probabilities and numerical values of their parameters are taken from \cite{CraciunBystable}. The association of variables  $q_i$ with the biochemical notations are: $E=q_1$, $S=q_2$, $ES=q_3$, $I=q_4$, $EI=q_5$ and $ESI=q_6$. We study the paths of the substrate molecule $S$ that show the bistable character of the biocircuit. The path of the $S$-molecule may move between a low and a high state. Because the behavior of $S$ is sufficient to show the bistability we decided to eliminate the product molecule $P$ from the Eq.(\ref{FEq:ESI}) and worked with 6 molecules instead of 7. We changed $ES \rightarrow E+P$ from \cite{CraciunBystable} with $T_3 = k_{prot}q_3 $ to $ES \rightarrow E$ with the same transition rate.

\begin{flalign}\label{table:ESIreactions}
  E&+S \rightarrow ES, \, T_1= k_{1}^{+} q_1 q_2, \; k_{1}^{+}=25979.537&\\
  E&S \rightarrow E+S, \, T_2 = k_1^{-}q_3,\;k_1^{-}=\,3.3722455\\
  E&S \rightarrow E, \, T_3 = k_{prot}q_3,\; k_{prot}=5844.999 \\
  E&+I \rightarrow EI, \, T_4= k_{2}^{+} q_1 q_4,\;k_{2}^{+}=5.3341555 \\
  E&I \rightarrow E+I, \, T_5= k_{2}^{-} q_5,\;k_{2}^{-}=16623.325  \\
  E&S+I \rightarrow ESI, \, T_6= k_{3}^{+} q_3 q_4,\; k_{3}^{+}=12200.836 \\
  E&SI \rightarrow ES+I, \, T_7= k_{3}^{-} q_6,\; k_{3}^{-}=1472.3849 \\
  E&SI \rightarrow EI+S, \, T_8= k_{4}^{-} q_6,\; k_{4}^{-}=15145.809\\
  E&I+S \rightarrow ESI, \, T_9= k_{4}^{+} q_2 q_5,\;k_{4}^{+}=9647.324\\\nonumber
\end{flalign}

Molecules $S$ and $I$ degrade proportional with their respective number.

\begin{eqnarray}
  S &\rightarrow& \emptyset, \quad T_{10}= \xi_{S} q_2,\, \xi_{S}=1\\
  I &\rightarrow& \emptyset, \quad T_{11}= \xi_{I} q_4,\,\xi_{I}=1
\end{eqnarray}

Molecules $S$ and $I$ accumulate, being coupled to external generators.

\begin{eqnarray}
   &\rightarrow& S, \quad T_{12}= G_S,\, G_S= 1734.2661 \\
   &\rightarrow& I, \quad T_{13}= G_I,\, G_I=1
\end{eqnarray}

The initial conditions for the Gillespie simulation at $t=0$ are $E=2$, $S=400$, $ES=0$, $I=1$, $EI=0$ and $ESI=0$. For the LC-method we take $0$ as $10^{-10}$ for the same reason given in Sec.\ref{SEquilibriumReaction} Suplemmental Material, and  so we used $E=2$, $S=400$, $ES=10^{-10}$, $I=1$, $EI=10^{-10}$ and $ESI=10^{-10}$. For the LC-method we need initial conditions for the second order moments. We used $F_{ii}(0)=\text{Abs}[q_i(q_i-1)]$ where the absolute value, Abs, was necessary only for the case of $10^{-10}$ initial condition. The initial values for the correlations were $F_{ij}(0)=q_iq_j$ for all $i<j$ with $i,j=1\dots6$. The molecules are considered to be uncorrelated at $t=0$, with an initial probability distribution $F(z,0)=z_1^2 z_2^{400}z_3^{0}z_4^{1}z_5^{0}z_6^{0}$. The differential equations were numerically solved on the interval $[0,4]$. The LC method provides the following master equation for the generating function $F(z,t)\equiv F(z_1,z_2,z_3,z_4,z_5,z_6,t)$.

\begin{equation}\label{FEq:ESI}
\begin{split}
 \partial_t F(z,t)=&\lambda_1 (z_1^{-1} z_2^{-1} z_3-1)k_1^{+} \frac{F_{12}(t)}{F_1(t)}z_1\partial_{z_1}F(z,t)+(1-\lambda_1) (z_1^{-1} z_2^{-1} z_3-1)k_1^{+} \frac{F_{12}(t)}{F_2(t)}z_2\partial_{z_2}F(z,t)+\\
 &\lambda_2 (z_1^{-1} z_4^{-1} z_5-1)k_2^{+} \frac{F_{14}(t)}{F_1(t)}z_1\partial_{z_1}F(z,t)+(1-\lambda_2) (z_1^{-1} z_4^{-1} z_5-1)k_2^{+} \frac{F_{14}(t)}{F_4(t)}z_4\partial_{z_4}F(z,t)+\\
 &\lambda_3 (z_3^{-1} z_4^{-1} z_6-1)k_3^{+} \frac{F_{34}(t)}{F_3(t)}z_3\partial_{z_3}F(z,t)+(1-\lambda_3) (z_3^{-1} z_4^{-1} z_6-1)k_3^{+} \frac{F_{34}(t)}{F_4(t)}z_4\partial_{z_4}F(z,t)+\\
 &\lambda_4 (z_2^{-1} z_5^{-1} z_6-1)k_4^{+} \frac{F_{25}(t)}{F_2(t)}z_2\partial_{z_2}F(z,t)+(1-\lambda_4) (z_2^{-1} z_5^{-1} z_6-1)k_4^{+} \frac{F_{25}(t)}{F_5(t)}z_5\partial_{z_5}F(z,t)+\\
 &(z_1 z_2 z_3^{-1}-1)k_1^{-}z_3\partial_{z_3}F(z,t)+(z_1 z_4 z_5^{-1}-1)k_2^{-}z_5\partial_{z_5}F(z,t)+
 (z_3 z_4 z_6^{-1}-1)k_3^{-}z_6\partial_{z_6}F(z,t)+\\
 &(z_2 z_5 z_6^{-1}-1)k_4^{-}z_6\partial_{z_6}F(z,t)+(z_1 z_3^{-1}-1) k_{prot} z_3 \partial_{z_3}F(z,t)+
 (z_2^{-1}-1) \xi_S z_2\partial_{z_2}F(z,t)+\\
 &(z_4^{-1}-1) \xi_I z_4\partial_{z_4}F(z,t)+(z_2-1) G_S F(z,t)+(z_4-1) G_I F(z,t)
 \end{split}
\end{equation}

\section{Ultrasensitive network}

The parameters used to simulate the network from Fig.\ref{fig:FigureMAPKV11.pdf} are as follows:

\begin{itemize}
\item The $\lambda$-parameters are $\lambda_1=1,\; \lambda_2=0.5,\; \lambda_3=0.5,\; \lambda_4=0.5,\; \lambda_5=0.5,\; \lambda_6=1,\; \lambda_7=0.75,\; \lambda_8=0.5,\; \lambda_9=0.25,\; \lambda_{10}=0$.
\item $a=10^{3.2}, d=10^{3}, k=10^3$. The LC-master equation contains 10 parts of the type described in Fig.\ref{fig:FigureMAPKV11.pdf} inset. For all 10 parts the  $a, d$ and $k$ parameters are equal. Equal parameters were used also in \cite{FerrellHuang}.
\item To avoid overcrowding Fig.\ref{fig:FigureMAPKV11.pdf} with names of each molecule we will use the following method to localize the molecules. With reference to the inset of Fig.\ref{fig:FigureMAPKV11.pdf}, the name of the intermediate molecule $z_3$ is constructed by concatenation of the name of $z_1$ and $z_2$, the concatenation symbol being the column $:$, $\text{Name}[z_3]=\text{Name}[z_1]:\text{Name}[z_2]$. The molecules $z_4$ and $z_5$ are obtained by dissociation of $z_3$. The name of $z_4$ is $\text{Name}[z_4]=\text{Name}[z_3]/\text{Name}[z_5]$, where / means that $z_5$ left the complex $z_3$ to obtain $z_4$. Similarly, $\text{Name}[z_5]=\text{Name}[z_3]/\text{Name}[z_4]$
\item To write the Master Equation in terms of $z_k$, $k=1\dots 22$, the molecules from Fig.\ref{fig:FigureMAPKV11.pdf} are denoted as follows:
        \begin{flalign}\label{table:MAPKmolecules}
        q_1&=E1,\; q_2=w1,\; q_3=\alpha,\; q_4=E2,\; q_5=\beta,\; &\\\nonumber
        q_6&=\tilde{\beta}=(\beta:E2)/E2,\; q_7=E3,\; q_8=w2,\; q_9=\gamma,\; \\\nonumber
        q_{10}&=\tilde{\gamma}=(\gamma:E3)/E3,\; q_{11}=w3,\; q_{12}=\delta,\; q_{13}=\alpha:E1\\\nonumber
        q_{14}&=E2:w1,\; q_{15}=\beta:E2,\; q_{16}=w2:((\beta:E2)/E2),\; \\\nonumber
        q_{17}&=E2:((\beta:E2)/E2),\; q_{18}=E3:w2,\; q_{19}=\gamma:E3,\; \\\nonumber
        q_{20}&=w3:((\gamma:E3)/E3),\; q_{21}=E3:((\gamma:E3)/E3),\; \\\nonumber
        q_{22}&=\delta:w3\\\nonumber
        \end{flalign}
\item The initial molecule numbers, at $t=0$ are
\begin{flalign}\label{table:MAPKmoleculesInitialValues}
    q_1&=1\dots30,\; q_2=30,\; q_3=300,\; q_4=10^{-15},\; &\\\nonumber
    q_5&=120,\; q_6=10^{-15},\; q_7=10^{-15},\; q_8=30,\; q_9=120, \\\nonumber
    q_{10}&=10^{-15},\; q_{11}=12,\; q_{12}=10^{-15},\; q_{13}=10^{-15}\\\nonumber
     q_{14}&=10^{-15},\; q_{15}=10^{-15},\; q_{16}=10^{-15},\; \\\nonumber
     q_{17}&=10^{-15},\; q_{18}=10^{-15},\; q_{19}=10^{-15},\; \\\nonumber
     q_{20}&=10^{-15},\; q_{21}=10^{-15},\; q_{22}=10^{-15}\\\nonumber
\end{flalign}
\end{itemize}

\begin{itemize}
\item The initial probability, at $t=0$ is taken to be $F=\prod_{i=1}^{22}z_i^{q_i}$
\end{itemize}

The relation to the MAPK notation from \cite{FerrellHuang} is:

\begin{table}[h]
\centering
\begin{tabular}{|c|c|c|}
  \hline
  $q_1=\text{E1}$ & $q_9 =  \text{MAPK}$ & $q_{17} =  \text{MAPKKPMAPKKKS}$ \\
  $q_2=\text{E2}$ & $q_{10} =  \text{MAPKP}$ & $q_{18} =  \text{MAPKKPPMAPKKPa}$ \\
  $q_3=\text{MAPKKK}$ & $q_{11} = \text{MAPKPa}$ & $q_{19} = \text{MAPKKPPMAPK}$ \\
  $q_4 = \text{MAPKKKS}$& $q_{12} =  \text{MAPKPP}$ & $q_{20} =  \text{MAPKPMAPKPa}$ \\
  $q_5 = \text{MAPKK}$ & $q_{13} =  \text{MAPKKKE1}$ & $q_{21} =  \text{MAPKPMAPKKPP}$ \\
  $q_6 = \text{MAPKKP}$ & $q_{14} =  \text{MAPKKKE2}$ & $q_{22} =  \text{MAPKKPPMAPKPa}$ \\
  $q_7 = \text{MAPKKPP}$ & $q_{15} =  \text{MAPKKMAPKKKS}$ &  \\
  $q_8 =  \text{MAPKKPa}$ & $q_{16} =  \text{MAPKKPMAPKKPa}$ &  \\
  \hline
\end{tabular}
\end{table}

\vspace{0.5cm}

The stochastic dynamics of the ultrasensitive network can be expressed in terms of the time-dependant Hill function: $F_{\delta}(t)=\frac{\omega_{\delta} t^{m_{\delta}}}{1+\alpha_{\delta} t^{m_{\delta}}}$ and $F_{\delta \delta}(t)=\frac{\omega_{\delta\delta} t^{m_{\delta\delta}}}{1+\alpha_{\delta\delta} t^{m_{\delta\delta}}}$. The parameters $m$, $\omega$ and $\alpha$ depend on the initial value $F_{E1}(0)$.  We used the time-dependant Hill functions to compute the response times $T_{1/2}$, $T_{1/2}^{-}$ and $T_{1/2}^{+}$.





\section{Reducing a network by splitting and projection}

The steps taken to obtain the equivalence of the pair ($E3$,$\delta$) from the ultrasensitive subnetwork network with the simplified network from Fig.\ref{fig:FigureMAPKV16.pdf}(b) were:

\begin{itemize}
\item The time interval over which the projection was computed was taken to be $[0,0.5]$ and was divided in $5000$ pieces. The functions  $F_{E3}(t)$, $F_{\delta}(t)$, $F_{E3\delta}(t)$, $F_{E3E3}(t)$, and $F_{\delta\delta}(t)$ were computed using the LC-method applied to the entire network of 22 molecules of Fig.\ref{fig:FigureMAPKV11.pdf}. Each function was sampled at $t_k=0.0001 (k-1)$ with $k=1\dots 5000$.
 \item For the simplified molecule network of Fig.\ref{fig:FigureMAPKV16.pdf}(b) the driving $\lambda$-parameter is $\lambda =0.5$ because of the identity of the molecules $1$ and $2$.
 \item The optimization procedure was carried in two steps. First the parameters $k_{+}$, $k_{-}$ were considered functions of time and a sequence of $k_{+}(t_k)$, $k_{-}(t_k)$ for each sampled time was obtained. This optimization gives an equivalent model for Fig.\ref{fig:FigureMAPKV16.pdf}(b) with time-dependent association and dissociation parameters.  For each time $t_k$ the unknowns  $k_{+}(t_k)$, $k_{-}(t_k)$, $G(t_k)$, $p(t_k)$, $n(t_k)$, $G_3(t_k)$, $p_3(t_k)$ and $n_3(t_k)$ were determined by minimizing the objective function:
 $(\frac{dF_{E3}}{dt}-\frac{dF_1}{dt})^2+(\frac{dF_{\delta}}{dt}-\frac{dF_3}{dt})^2+(\frac{dF_{E3E3}}{dt}-\frac{dF_{11}}{dt})^2+(\frac{dF_{\delta\delta}}{dt}-\frac{dF_{33}}{dt})^2+(\frac{dF_{E3\delta}}{dt}-\frac{dF_{13}}{dt})^2$
  computed at $t_k$. There is no need to include in the objective function the moments of the molecule labeled $2$ in Fig.\ref{fig:FigureMAPKV16.pdf}(b) because the time evolution  of this molecule is identical with the evolution of molecule $1$. The minimization was carried out through Mathematica command NMinimize with the DifferentialEvolution method, \cite{Wolfram}. The optimization constrain imposes that all the unknowns should be nonnegative.
 \item For the second optimization procedure we computed  the median value for the association and dissociation time-dependent parameters obtained from the first optimization: $k_{+}=0.00013$ and $k_{-}= 0.009$. This values were used for a second run of the optimization algorithm for which $k_{+}$ and $k_{-}$ are now known constants.
\end{itemize}



\begin{figure}[t]
\includegraphics[scale=0.5]{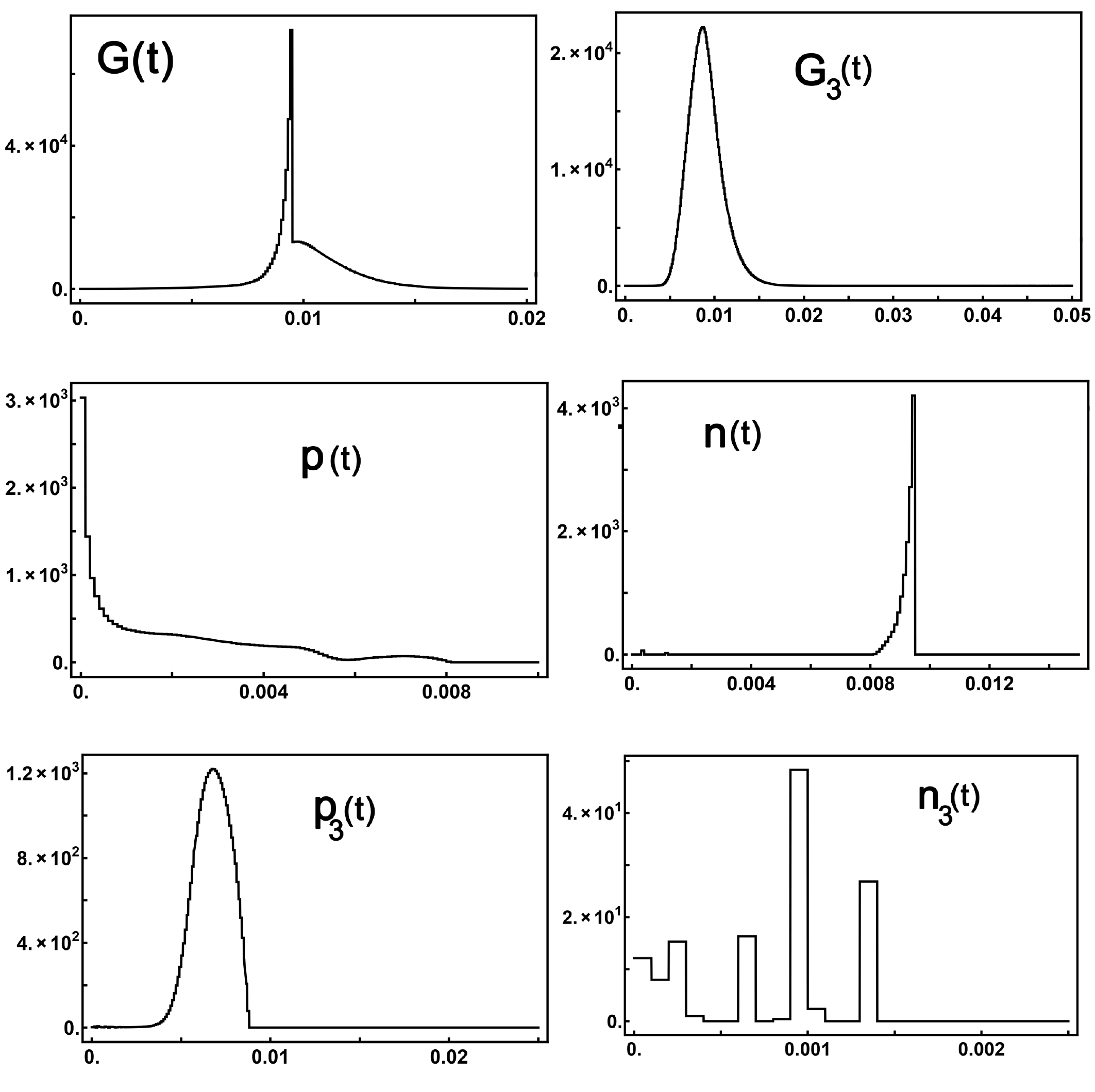}
\caption{\label{fig:FigureGeneratorsTwoMoleculeEquivalentSystemFlattened}The results of the second optimization for which $k_{+}=0.00013$ and $k_{-}= 0.009$. The horizontal axes represents time. The time horizon on which the systems from Fig.\ref{fig:FigureMAPKV16.pdf} were studied is $0.05$. Only the nonzero time interval on which the  generators $G$, $p$, $n$ and $G_3$, $p_3$, $n_3$ act was plotted.}
\end{figure}

\end{document}